\documentclass[aps,reprint,prb,showpacs,showkeys,preprintnumbers,amsmath,amssymb,superscriptaddress,floatfix]{revtex4-1}

\usepackage{graphicx}
\usepackage{hyperref}
\usepackage{dcolumn}
\usepackage{bm}
\usepackage[latin1]{inputenc}

\usepackage{color}
\usepackage{pstricks}

\newcounter{saveeqn}

\begin{document}

\title{Effects of strain on the valence band structure and exciton-polariton energies in ZnO}

\author{Markus R. Wagner}
\email{markus.wagner@tu-berlin.de}
\affiliation{Institute of solid state physics, Technische Universit\"at Berlin, 10623 Berlin, Germany}
\affiliation{ICN2 - Institut Catala de Nanociencia i Nanotecnologia, Campus UAB, 08193 Bellaterra (Barcelona), Spain}
\author{Gordon Callsen}
\affiliation{Institute of solid state physics, Technische Universit\"at Berlin, 10623 Berlin, Germany}
\author{Juan S. Reparaz}
\affiliation{Institute of solid state physics, Technische Universit\"at Berlin, 10623 Berlin, Germany}
\affiliation{ICN2 - Institut Catala de Nanociencia i Nanotecnologia, Campus UAB, 08193 Bellaterra (Barcelona), Spain}
\author{Ronny Kirste}
\affiliation{Institute of solid state physics, Technische Universit\"at Berlin, 10623 Berlin, Germany}
\affiliation{Department of Material Science and Engineering, NCSU, Raleigh, NC 27695, USA}
\author{Axel Hoffmann}
\affiliation{Institute of solid state physics, Technische Universit\"at Berlin, 10623 Berlin, Germany}
\author{Anna V. Rodina}
\affiliation{A. F. Ioffe Physico-Technical Institute, 194021 St. Petersburg, Russia}
\author{Andr{\'e} Schleife}
\email{a.schleife@llnl.gov}
\affiliation{Condensed Matter and Materials Division, Lawrence Livermore National Laboratory, Livermore, CA 94550, USA}
\affiliation{Institut f\"ur Festk\"orpertheorie und -optik, Friedrich-Schiller-Universit\"at, 07743 Jena, Germany}
\affiliation{European Theoretical Spectroscopy Facility (ETSF)}
\author{Friedhelm Bechstedt}
\affiliation{Institut f\"ur Festk\"orpertheorie und -optik, Friedrich-Schiller-Universit\"at, 07743 Jena, Germany}
\affiliation{European Theoretical Spectroscopy Facility (ETSF)}
\author{Matthew R. Phillips}
\affiliation{Department of Physics and Advanced Materials, University of Technology Sydney, NSW 2007, Australia}
\date{\today}

\begin{abstract}
The uniaxial stress dependence of the band structure and the exciton-polariton transitions in wurtzite ZnO is thoroughly studied using modern first-principles calculations based on the HSE+$G_0W_0$ approach, $\bf k \cdot \bf p$ modeling using the deformation potential framework, and polarized photoluminescence measurements. The ordering of the valence bands [A($\Gamma_7$), B($\Gamma_9$), C($\Gamma_7$)] is found to be robust even for high uniaxial and biaxial strains.  Theoretical results for the uniaxial pressure coefficients and splitting rates of the A, B, and C valence bands and their optical transitions are obtained including the effects of the spin-orbit interaction. The excitonic deformation potentials are derived and the stress rates for hydrostatic pressure are determined based on the results for uniaxial and biaxial stress. In addition, the theory for the stress dependence of the exchange interaction and longitudinal-transversal splitting of the exciton-polaritons is developed using the basic exciton functions of the quasi-cubic approximation and taking the interaction between all exciton states into account. It is shown that the consideration of these effects is crucial for an accurate description of the stress dependence of the optical spectra in ZnO. The theoretical results are compared to polarized photoluminescence measurements of different ZnO substrates as function of uniaxial pressure and experimental values reported in the literature demonstrating an excellent agreement with the computed pressure coefficients.\\
DOI: 10.1103/PhysRevB.88.235210
\end{abstract}


\maketitle

\section{Introduction}

The electronic band structure and the related exciton-polariton transitions in wurtzite semiconductors were investigated in countless experimental and theoretical studies over the past five to six decades. Indeed for ZnO important contributions in the field were already published in the early 1960s \cite{hopfield59, hopfield60b, thomas60, reynolds65, park66}. Nevertheless, fundamental properties such as the ordering of the valence bands (VBs) and the influence of stress on the exciton-polariton transitions remain controversial to the present day. Magnetic fields and external pressure constitute powerful tools to obtain detailed information about electronic band structures and optical transitions by studying the exciton-polariton fine structure. In addition, in recent years, significant advances in computational power and theoretical algorithms have enabled electronic-structure calculations, including quasiparticle (QP) effects and electron-hole interaction which reproduce experimental results to a high degree of precision. Consequently, it is now possible to conduct complex calculations taking into account, for instance, the influence of stress on the electronic band structure and the exciton-polariton transitions in direct comparison to experimental results.

The long-standing disagreement of the valence band ordering in ZnO is one important example for the necessity to comprehend the effects of built-in strain and external stress on the electronic band structure and optical transitions in detail. Recent magneto-optical measurements of free and bound excitons provide strong evidence that the topmost A valence band has $\Gamma_7$ symmetry~\cite{wagner09magneto, rodina04}. These results are in accordance with \emph{first-principles} QP band-structure calculations \cite{Schleife:2007, lambrecht02} as well as a multitude of earlier theoretical and experimental works supporting the valence band ordering originally proposed by Thomas in 1960.\cite{thomas60} By contrast, several publications exist which postulate $\Gamma_9$ symmetry for the A valence band (for a summary of the relevant literature see e.g.\ Refs.\ \onlinecite{wagner09magneto} and \onlinecite{ZnOBook}). While some of the conflicting results have been resolved\cite{reynolds99, rodina04}, the important question remains if the presence of strain (or the application of external stress) that preserves the wurtzite structure of ZnO (i.e.\ uniaxial pressure along the $\bm c$ axis, biaxial stress perpendicular to the $\bm c$ axis, and hydrostatic pressure) may result in a reversal of the valence band ordering and thus could explain the different assignments in the literature.

Such a reversal of the A and B VB symmetry as function of strain was discussed e.g.\ for the wurtzite III-V semiconductors GaN~\cite{gil96, shikanai97, fu09} and AlN\cite{ikeda07}. Gil \emph{et al.}~\cite{gil96} first predicted a crossing of the upper two VBs in GaN for biaxial in-plane strain. This prediction was supported by Shikanai \emph{et al.}\cite{shikanai97} and more recently by Fu \emph{et al.}\cite{fu09} who reported a reversal of the VB ordering for uniaxial strain values of $\varepsilon_{zz}\approx -0.07\%$ and $\varepsilon_{zz}=-0.034\%$, respectively. Ikeda \emph{et al.}\cite{ikeda07} and Fu \emph{et al.}\cite{fu09apl} calculated an exchange of the upper two VB characteristics in AlN for uniaxial strain values of $\varepsilon_{zz}>0.70\%$ and $\varepsilon_{zz}>0.98\%$, respectively. 

In the case of ZnO, Gil \emph{et al.} suggested a reversal of the A and B valence band ordering for a biaxial compressive stress of $P_b\approx 2$~kBar which corresponds to a uniaxial part of the strain tensor of $\varepsilon_{zz}\approx 0.9\cdot 10^{-3}$ and thus concluded that the VB ordering in ZnO is quite sensitive to strain.\cite{gil01jjap} However, the authors had to rely on exciton energies reported in the literature which included only one work by Butkhuzi \emph{et al.}\cite{butkhuzi98} with sufficiently shifted free exciton energies to suggest a large in-plain strain and a strain-related reversal of the VB ordering. A closer look into this work reveals several major problems which render the reported transition energies in Ref.\ \onlinecite{butkhuzi98} questionable for a strain analysis: (i) The luminescence spectra were acquired at 77~K instead of 4~K resulting in a shift of the transition energies in accordance with the temperature dependence of the band gap~\cite{paessler03}, (ii) the luminescence was excited by a pulsed nitrogen laser with high pulse energy leading to an excitation density related shift of the observed exciton lines\cite{reynolds00a}, and (iii) an inaccurate proportionality factor was used to convert the wavelength values (nm) into energy (eV) resulting in energy values which are too small by about 2.5~meV. The combination of these effects leads to significant deviations of the exciton energies and consequently results in misleading strain values and conclusions in the experimental part of Ref.\ \onlinecite{gil01jjap}. This situation provides a strong motivation to revisit the stress dependence of the valence band ordering in ZnO in detail. The large research interest in this field is also reflected by several most recent publications about the strain dependence of the electronics bands in related materials.\cite{yadav12, wu12, yan12}

Apart from the influence of stress and strain on the VB ordering, important elastic and electronic parameters such as the phonon- and exciton deformation potentials (DP) can be derived by e.g.\ Raman and luminescence studies as function of applied pressure. Hydrostatic pressure was widely used to study the phase transition from the wurtzite to the rocksalt structure\cite{karzel96, desgreniers98} as well as a variety of phonon related parameters such as the Gr\"uneisen parameters and the pressure dependence of the Born effective charge.\cite{decremps02, reparaz10pressure} In addition, the phonon deformation potentials were determined by Raman spectroscopy as function of uniaxial pressure.\cite{callsen11apl} However, only few sources are available for the electronic deformation potentials. For GaN and AlN these were recently studied using reflectance spectroscopy under uniaxial stress by Ishii et al.\cite{ishii10, ishii13} In the case of ZnO, experimental studies and theoretical calculations were published without including the effects of spin-orbit coupling (SOC) \cite{langer70, wrzesinski97, yan12}. The effects of uniaxial pressure were experimentally studied in great detail by Wrzesinski and Fr\"ohlich using two-photon and three-photon spectroscopy but the spin-orbit splitting and the contributions of exchange interaction were not resolved and discussed separately.\cite{wrzesinski97phd, wrzesinski97, wrzesinski98jcg}

In order to remedy this situation, we investigate in this work the effects of stress on the band structure and exciton-polariton energies in wurtzite ZnO both theoretically and experimentally. We perform a detailed theoretical analysis of the effects of external uniaxial stress, biaxial stress, and hydrostatic pressure on the A-B and A-C exciton-polariton splittings, the exchange splitting, and the longitudinal-transversal (LT) splitting using the quasi-cubic approximation and taking the interaction between all exciton states into account.\cite{wrzesinski97, wrzesinski98ssc, wrzesinski98jcg} From the results of \emph{first-principles} calculations based on the HSE+$G_0W_0$ method (see section \ref{abinitio}), we compute the deformation potentials that describe all effects of external stress which preserve the wurtzite symmetry of ZnO. The full set of the DP constants includes those describing the spin-dependent part (allowing for the anisotropy of the spin-orbit interaction) as well as the kinetic-energy part of the effective Hamiltonian in the quasi-cubic approximation. Special attention is paid to the effects of stress on the spin-orbit interaction constant and its anisotropy. The results of these considerations are compared with uniaxial stress-dependent photoluminescence (PL) measurements for ZnO substrates from different manufacturers as well as reported data in the literature.

The paper is structured as follows: After the description of the theoretical and experimental techniques in sections~\ref{abinitio} and \ref{setup}, we first present the results of the \emph{ab-initio} calculations for the electronic bands as function of pure uniaxial and biaxial stress and their combination (section~\ref{stress}). The excitonic deformation potentials including the stress dependence of the spin-orbit interaction are discussed in section~\ref{defpot}. The theoretical considerations are complemented by measurements of the exciton polariton energies as function of uniaxial stress (section~\ref{results}). Finally, the experimentally resolved exciton-polariton fine structure is used to investigate the effects of stress on the exchange interaction in section~\ref{pol} and the results are summarized in section \ref{conclusion}.

\section{\label{abinitio}Computational approach}

In order to compute energies of electronic states that can be compared to experimental results, it is important to account for single-particle effects which are computed by solving a QP equation that properly accounts for the electronic self energy $\Sigma$. Nowadays, Hedin's $GW$ approximation \cite{Hedin:1965} of $\Sigma$ is the state-of-the-art approach \cite{Aulbur:1999,Onida:2002}. Due to the high computational cost of a self-consistent solution of the QP equation, first-order perturbation theory is used to obtain QP energies by correcting an initial electronic structure \cite{Hybertsen:1986}. This so-called $G_0W_0$ approach requires an initial electronic structure not too far from the final results. We rely on solving a generalized Kohn-Sham equation \cite{Kohn:1965,Seidl:1996}, since for ZnO it has been shown before that the nonlocal HSE hybrid functional\cite{Heyd:2006,Krukau:2006,Paier:2006} using a parameter of $\omega=0.15$ a.u.$^{-1}$ (see Ref.\ \onlinecite{Paier:2006_b} for details) provides an excellent starting point for $G_0W_0$ calculations \cite{Schleife:2009,Schleife:2008,Schleife:2009_c,King:2009_b,Preston:2008}.

Furthermore, in the present work the influence of the spin-orbit interaction on electronic states of strained ZnO is investigated. Earlier studies \cite{Schleife:2007,Schleife:2008,Schleife:2009_c} indicate that the SOC only slightly modifies the Kohn-Sham states of ZnO; the impact on the screened Coulomb potential $W$ is expected to be small. Therefore, the influence of the spin-orbit interaction on the QP \emph{corrections} is neglected here. QP energies that account for the SOC-related effects are calculated by adding QP shifts calculated without SOC to HSE eigenvalues that include spin-orbit-interaction effects \cite{Schleife:2007,Schleife:2009_c}. This approach is expected to produce reasonable results because the absolute spin-orbit-induced shifts are small compared to the QP corrections for ZnO.

The QP electronic structures (including SOC) of the uniaxially and biaxially strained cells are calculated using the Vienna \emph{Ab-Initio} Simulation Package
(VASP) \cite{Kresse:1996,Kresse:1996_b,Shishkin:2006}. The O\,$2s$ and O\,$2p$ electrons as well as the Zn\,$3d$ and Zn\,$4s$ states are treated as valence electrons. The projector-augmented wave method is applied to describe the wave functions in the core regions \cite{Bloechl:1994,Kresse:1999,Hobbs:2000}. A plane-wave cutoff energy of $400$~eV is chosen to obtain converged results. In order to carry out the HSE+$G_0W_0$ calculations, $\Gamma$-centered $8\times 8\times 6$ Monkhorst-Pack\cite{Monkhorst:1976} ${\bf k}$-point meshs along with at least 300 conduction bands were used in this work. We are aware that this method slightly underestimates the fundamental gaps, however, it is well suited for the description of the strain dependence of the QP electronic structure.

\section{\label{setup}Experimental setup}

\begin{figure}
\begin{center}
\includegraphics[width=6cm, keepaspectratio]{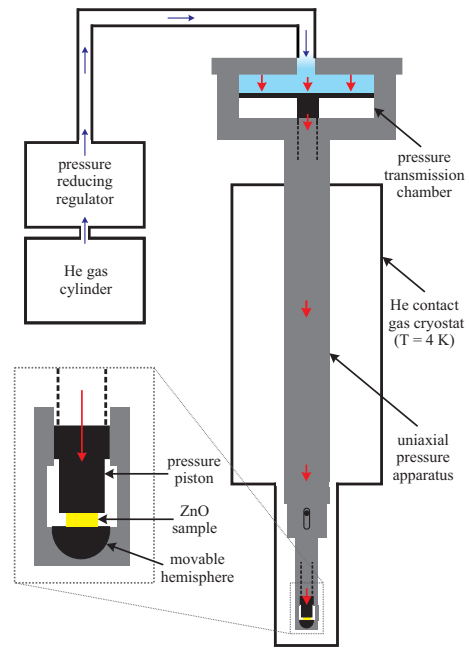}
\caption{\label{uniaxial-apparatus}(Color online) Illustration of the pressure apparatus used for the uniaxial pressure measurements. The sample is placed between a movable hardened steel hemisphere and the lower end of the pressure piston (enlarged section). Pressure is applied by an external pressurized He gas bottle.}
\end{center}
\end{figure}

The theoretical calculations are complemented by experimental results: PL measurements of the free exciton-polaritons as function of uniaxial pressure are performed for ZnO substrates from different suppliers (CrysTec, Tokyo Denpa, and UniWafer). The luminescence of the samples is excited by the 325~nm line of a HeCd laser with an excitation power of 20~mW and detected using a bi-alkali photomultiplier attached to a SPEX 1~m double monochromator with a spectral resolution of 50~$\mu$eV. Uniaxial pressure measurements were performed in a self-built pressure apparatus (Fig.~\ref{uniaxial-apparatus}) which is placed inside a helium bath cryostat with a temperature of 4~K. The pressure on the samples was applied via a stainless-steel piston using a pneumatic construction. The primary gas pressure in the pressure transmitting chamber on top of the piston was fine tuned by a two stage pressure regulator with an error of 5~mbar. The pressure in this chamber was used to push the piston downwards against a hardened steel hemisphere in between which the sample was placed. The surface ratio between the pressure-transmitting chamber and the surface area of the samples determines the pressure transmission ratio. Typical values are between 500 and 1000 for samples with surface areas between 2 to 4~mm$^2$. Assuming a precision of $\pm10$ $\mu$m for the size determination of the samples and the pressure stamp and considering the pressure accuracy in the primary gas chamber, the error of the applied pressure is estimated to $\pm0.002$ GPa. For the pressure-dependent measurements the $\bm c$ axis was oriented parallel to the direction of the uniaxial pressure ($\bm{P} \| \bm{c}$). The luminescence was excited and detected from the edge of the sample in $\bm{k} \bot \bm{c}$ direction with $\bm{k}$ indicating the direction of the exciting and emitted light. Polarization-dependent measurements were performed using a linear polarizer to select the $\bm{E} \| \bm{c}$ and $\bm{E} \bot \bm{c}$ components of the luminescence.

\section{\label{stress}Strain dependence of electronic bands}

The strain dependence of the conduction and valence band extrema is theoretically determined by \emph{ab-initio} calculations as described in section \ref{abinitio}. The lattice of unstrained ZnO is characterized by the equilibrium values of the lattice constants $a_0$ and $c_0$. In the present work we use the theoretical values reported by Schleife et al.\cite{Schleife:2006} for the unstrained values of the lattice constants. As perturbation we consider uniaxial stress along the $\bm c$ axis, isotropic biaxial in-plane stress, and their simultaneous combination which preserve the wurtzite symmetry of ZnO. The following subsections briefly explain the procedure for these three cases.

\subsection{\label{uniax}Pure uniaxial and pure biaxial stress}

Pure uniaxial stress $P_u$ is defined by vanishing forces in the plane and constant forces along the $\bm c$ axis ($\bm z$ axis) whereas pure biaxial stress $P_b$ is defined by constant forces in the plane and vanishing forces along the $\bm c$ axis: \cite{Wagner02}
\begin{align}
\sigma_{xx}=\sigma_{yy}=0, \sigma_{zz}=-P_u,\\
\sigma_{xx}=\sigma_{yy}=-P_{b}, \sigma_{zz}=0,
\end{align}
with $P_u$ ($P_b$) being positive in the case of uniaxial (biaxial) compression along the $\bm z$ axis (in the plane) and negative otherwise.

Using Hooke's law with the condition $\sigma_{xx}=0$ or $\sigma_{zz}=0$, one obtains for the cases of uniaxial and biaxial stress
\begin{align}
\varepsilon_{xx}= -\nu \varepsilon_{zz}=- \nu  \varepsilon_{u},\\
\varepsilon_{zz}=- R^b \varepsilon_{xx}=- R^b \varepsilon_{b},
\end{align}
where $\nu$ is the Poisson ratio and $R^b$ is the biaxial relaxation coefficient.

Within the \emph{ab-initio} calculations, we vary the lattice constants $a$ and $c$ from their relaxed value $a_0$ and $c_0$ to the uniaxially or biaxially stressed value $c_u$ or $a_b$ in order to obtain the value $a_u$ or $c_b$, corresponding to the uniaxial and biaxial stress, respectively. For each value of the uniaxial strain $\varepsilon_{u}=(c_u-c_0)/c_0$ or biaxial strain $\varepsilon_{b}=(a_b-a_0)/a_0$, we then compute the QP energies (including spin-orbit interaction, cf.\ Sec.\ \ref{abinitio}) of the conduction band and valence band extrema. We compute linear fits to these results so that the QP energies $E_i$ can be expressed as 
\begin{align}
\label{eq:euniax}
E_u^\varepsilon = &E^0+d_{u} \varepsilon_{u} = E^0-\frac{d_u}{E} P_{u},\\
E_b^\varepsilon = &E^0+d_{b} \varepsilon_{b} = E^0-\frac{d_b}{Y} P_{b},
\end{align}
where $d_{u}$ and $d_{b}$ are the uniaxial and biaxial strain coefficients, respectively. These strain coefficient are summarized for the conduction and valence bands with and without SOC in Table\ \ref{EABC}.

The uniaxial stress $P_u$ and biaxial stress $P_b$ are then given by
\begin{align}
P_u = &-E\varepsilon_u = -E(c_u-c_0)/c_0,\\
P_b = &-Y\varepsilon_b = -Y(a_b-a_0)/a_0,
\end{align}
with the Young modulus E and biaxial modulus Y.

\begin{figure}
\begin{center}
\includegraphics[width=\columnwidth, keepaspectratio]{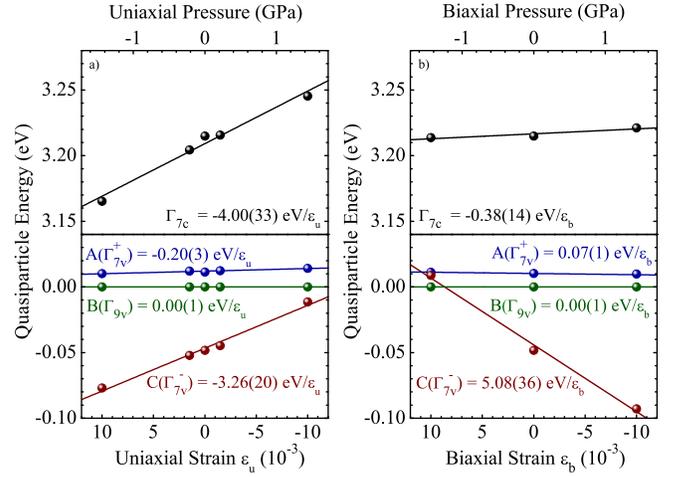}
\caption{\label{theory}(Color online) Results of \emph{ab-initio} calculations for the QP energies (including spin-orbit coupling) of the conduction- and valence-band states at the $\Gamma$ point as a function of uniaxial strain along the $\bm c$ axis (a) and biaxial in-plane strain (b). Solid lines represent linear fits to the data. The results of these fits are listed in Table\ \ref{EABC}.}
\end{center}
\end{figure}

The computed values for the QP energies of the conduction band and valence bands as function of uniaxial and biaxial strain are visualized in Fig.~\ref{theory} together with the linear fits to these results. The upper A and B bands (without strain) in Fig.~\ref{theory} have $\Gamma_7$ and $\Gamma_9$ symmetry, respectively, and originate from the $\Gamma_{5v}$ band (without considering SOC). This ordering indicates a negative SOC in ZnO, in agreement with earlier \emph{ab-initio} calculations\cite{lambrecht02} and the conclusions based on magneto-optical studies by Wagner \emph{et al.}\cite{wagner09magneto} and Rodina \emph{et al.}\cite{rodina04} One can note that the shift of the A and B valence bands as function of uniaxial and biaxial strain is very small. Consequently, the strain dependence of the energy splitting between these two bands is also small which clearly shows that the symmetry ordering of the A and B valence bands cannot be reversed for any reasonable tensile or compressive strain. Assuming a linear relation even at very high strain values, a uniaxial tensile strain of $\varepsilon_{u}=0.058$ or a biaxial compressive strain of $\varepsilon_{b}=-0.166$ would be required to result in a reversal of the topmost two VBs. Apparently, this amount of strain is neither realistic for uniaxial or biaxial pressure experiments nor for strain due to strongly lattice mismatched hetero-epitaxial growth. 

\begin{table*}
\caption{\label{EABC}Energy splittings, strain rates, and stress rates for wurtzite ZnO. Energy splittings $E=E^0+d_u\cdot\varepsilon_u+d_b\cdot\varepsilon_b$ with relaxed energy values $E^0$ as well as uniaxial $d_u$ and biaxial $d_b$ strain rates are obtained from \emph{ab-initio} calculations and fitting of the ${\bf k}\cdot{\bf p}$ band-structure Hamiltonian in comparison to experimental results. Uniaxial, biaxial, and hydrostatic stress rates are given for the energy splittings and ${\bf k}\cdot{\bf p}$ parameters in comparison to experimental results for uniaxial pressure (this work) and hydrostatic pressure (Ref.\ \onlinecite{mang95}). The stress rates are calculated using the elastic moduli $Y$ and $E$ from Ref.\ \onlinecite{Schleife:2007}.}
\begin{ruledtabular}
\begin{tabular}{lcccccccccc}
& ${\bf k}\cdot{\bf p}$ & $E^0$ & $E^0$ & $d_u$ & $d_b$ & $dE/dP_u$ 	& $dE/dP_u$   & $dE/dP_b$ & $dE/dP_h$ &  $dE/dP_h$ \\
&											 &  calc.	&	exp.$^a$ & calc.	&	calc. & calc. 		& exp.$^a$ 	 &  calc. 	&  calc.  	& 	 exp.$^b$ \\
& & (meV) & (meV) & (meV) & (meV) & \multicolumn{2}{c}{(meV/GPa)} & \multicolumn{2}{c}{(meV/GPa)} & {(meV/GPa)}\\ 
\hline
$\Gamma_{1c}-\Gamma_{5v}$ & $E_{\rm g}$ 								& 3205 	&	- & $-3960$ & $-370$  & 27.6 		& -- 			& 	1.7 & 29.3 &  -- \\
$\Gamma_{5v}-\Gamma_{1v}$   & $\Delta_{\rm cr}$ 				& 51.0	& - &  3180 & $-5080$ & $-22.1$ 	& -- 		& 23.6  & 1.5  & -- \\
\hline
$\Gamma_{7c}-\Gamma_{7v}^{+}$	& $E_{\rm g}(\mathrm{A})$ & 3200  & 3312.5 & $-3800$ 	& $-450$ & 26.4  		& 23.4 & 2.1 		& 28.5 	&  24.7\\ 
$\Gamma_{7c}-\Gamma_{9v}$	& $E_{\rm g}(\mathrm{B})$ 		& 3212 	& 3319.5 & $-4000$ 	& $-380$ & 27.8 			& 24.7 & 1.7 		& 29.5 	&  25.3\\	
$\Gamma_{7c}-\Gamma_{7v}^{-}$ & $E_{\rm g}(\mathrm{C})$ & 3258	& 3358.5 & $-740$		&	$-5460$ &  5.2  		& 5.8  & 25.3 	& 30.5 	&  26.8\\ 
$\Gamma_{7v}^{+}-\Gamma_{9v}$  & $E_{\rm AB}$						& 11.6 	& 7.0 	 &	$-200$ 	& 70   & 1.4  			& 1.3 & $-0.3$ 	& 1.1 	&  0.6\\ 
$\Gamma_{9v}-\Gamma_{7v}^{-}$  & $E_{\rm BC}$						& 45.7 	& 39.0   & 3260 		& $-5080$ & $-22.7$ 	& $-18.9$ & 23.6 & 0.9 	&  1.5\\ 
$\Gamma_{7v}^{+}-\Gamma_{7v}^{-}$ & $E_{\rm AC}$				& 57.3 	& 46.0   & 3060 		& $-5015$ & $-21.3$ 	& $-17.6$ & 23.3 & 2.0 	&  2.1\\ 
\hline
 & $\Delta_{\rm so}^\|$ 																& $-16.9$ & - 	 &   285 & $-75$  & $-2.0$ 		& -- & 0.4 			& $-1.6$&  --\\
 & $\Delta_{\rm so}^\bot$ 															& $-9.1$  & - 	 &	 $-110$ & 140 & 0.8	 		& -- & $-0.7$ 	& 0.1 	&  --\\
\end{tabular}
\end{ruledtabular}
$^a$ Experimental values are extracted from the experimental data for the exciton-polariton zero-stress energies and stress rates obtained in Sec.\ \ref{results} and using the ${\bm k}\cdot{\bm p}$ model as described in Sec.\ \ref{pol}.\\
$^b$ Experimental values of hydrostatic pressure measurements reported by Mang et al. in Ref.\ \onlinecite{mang95}.
\end{table*}

In contrast to the weak strain dependence of the A and B valence bands, the slope of the C valence band energy maximum as function of strain is several times larger. For a sufficiently large uniaxial compressive strain of $\varepsilon_{u}=-0.014$ or biaxial tensile strain of $\varepsilon_{b}=0.009$, the C($\Gamma_7$) valence band ($\Gamma_{1v}$ without SOC) will cross the B($\Gamma_9$) valence band and even become the topmost VB for a uniaxial strain of $\varepsilon_{u}=-0.019$ or biaxial tensile strain of $\varepsilon_{b}=0.011$, respectively. This behavior reflects the significantly stronger influence of stress on the crystal field splitting compared to the SOC as also shown by the absolute values of the uniaxial and biaxial strain rates ($d_u$ and $d_b$) in Table\ \ref{EABC}. It should be noted that these values only originate from linear extrapolations of the strain rates. In reality, anti-crossing of valence bands with equal symmetry and symmetry mixing of VB states will lead to a more complex non-linear behavior in the high strain regime.\cite{lambrecht02} In any case, the A valence band would maintain $\Gamma_7$ symmetry for any practically relevant strain values, thus excluding the possibility of a reversed symmetry ordering of the valence bands in ZnO by uniaxial or biaxial strain.

In order to facilitate the direct comparison between \emph{ab-initio} calculations and experimental results, we anticipate the results of Sec.\ \ref{results} and Sec.\ \ref{pol} and list the experimental values for the zero-stress energy splittings between the electronic bands in the forth column ($E^0$) of Table\ \ref{EABC}. These values are obtained from the experimentally observed zero-stress position of the exciton-polariton transitions in Sec.\ \ref{results} and using the ${\bm k}\cdot{\bm p}$ model as described in Sec.\ \ref{pol}. The results in Table~\ref{tZeroParameter} are than subtracted by the exciton binding energies as determined in Sec.\ \ref{results}. The resulting discrepancy for the zero-stress energy values of the band gap energies $E^0$ between our \emph{ab-initio} calculations and our experimental results is found to be smaller than 3.5\%. The small deviation in bandgap energies expresses the nowadays achievable precision of the theoretical approach even for the typically difficult prediction of wide-bandgap materials and is very encouraging for a precise theoretical prediction of the stress rates in the next section.

\subsection{\label{comb}Combination of uniaxial and biaxial stress: linear deformation potential approach}

Following the separate treatment of uniaxial and biaxial stress, let us consider the situation when both uniaxial stress $P_u$ and biaxial stress $P_b$ are applied simultaneously:
\begin{align}
\label{s1}
\sigma_{xx}=\sigma_{yy}=&-P_b, \\
\label{s2}
\sigma_{zz}=&-P_u.
\end{align}
The components of the strain tensor are obtained from the components of the stress tensor using Hooke's law with 
\begin{align}
\label{e1}
\varepsilon_{xx}=\varepsilon_{yy}=& \nu\frac{P_u}{E} - \frac{P_b}{Y} = -\nu \varepsilon_{u} + \varepsilon_{b}, \\
\label{e2}
\varepsilon_{zz}=& -\frac{P_u}{E} + R^b\frac{P_b}{Y} = \varepsilon_{u}- R^b \varepsilon_{b}.
\end{align}
This shows that the combination of uniaxial and biaxial stress conserves the hexagonal symmetry of the strain tensor which is a linear combination of the uniaxial and biaxial strain components. In the linear approximation of the ${\bf k}\cdot{\bf p}$ deformation potential (DP) model any characteristic energy $E^\varepsilon$ depends on the strain tensor via
\begin{equation}
\label{def}
E^\varepsilon=E^0 +  D_i \varepsilon_{zz} + D_j
(\varepsilon_{xx}+\varepsilon_{yy}),
\end{equation}
where $D_i$ and $D_j$ are the respective DP constants\cite{bir74}. The stress and strain dependences for the combined case are derived by inserting the strain tensors of Eqs.\ \eqref{e1} and \eqref{e2} into Eq.\ \eqref{def}:
\begin{align}
\label{bu}
E^\varepsilon&=E^0+d_u \varepsilon_u + d_b \varepsilon_b = E^0-d_u\frac{P_u}{E} -d_b \frac{P_b}{Y}, \\
\label{du}
d_u&=D_i- 2\nu D_j,\\
\label{db}
d_b&=2D_j-R_bD_i.
\end{align}

The characteristic energies depend linearly on the combination of the uniaxial and biaxial stress $P_u$ and $P_b$. The resulting stress rates of the band gap energies and the VB splittings are listed in Table\ \ref{EABC} for uniaxial pressure ($dE/dP_u$) and biaxial pressure ($dE/dP_b$). While the \emph{ab-initio} calculations were performed exclusively for pure uniaxial or pure biaxial stress, any combination of the two can be taken into account to compute the QP energies using Eq.\ \eqref{bu}.

For the special case of hydrostatic pressure $P_h=P_u=P_b$, the stress rates $dE/dP_h=dE/dP_u+dE/dP_b$ are also listed in Table~\ref{EABC} and are compared to hydrostatic pressure measurements \cite{mang95}. The computed hydrostatic pressure rate of 28.5 meV/GPa for the band gap energy compares well with previously published experimental values between 23.5 meV/GPa and 29.7 meV/GPa (see Ref.\ \onlinecite{chen06} and references therein). In addition, the stress rates for the different VB splittings are in very good agreement with the experimental values reported by Mang \emph{et al.}\cite{mang95} (cf.\ Table\ \ref{EABC}). This agreement between our calculations and hydrostatic pressure measurements is encouraging for achieving also a good theoretical description of the uniaxial pressure measurements in this work. However, before we turn to the results of those measurements in Sec.\ \ref{results}, we introduce the excitonic DPs and discuss
the stress dependence of the SOC in the following section.

\section{\label{defpot}Excitonic deformation potential constants}

We start with the Hamiltonian of the relative electron-hole motion of an exciton for a hexagonal semiconductor \cite{lambrecht02,rodina01}: 
\begin{align}
\begin{split}
\label{exciton}
{\hat H}_{\rm ex}({\bm k})={\hat H}_{e}({\bm k})&-{\hat H}_{h}(-{\bm k})\\
&-\frac{e^2}{\sqrt{\epsilon_\|\epsilon_\bot(x^2+y^2)+\epsilon_\bot^{ 2}z^2}},
\end{split}
\end{align}
where ${\bm k} = -i{\bm \nabla}$ is the wave vector of the relative motion, $(x,y,z)={\bm r}$ are the relative electron-hole coordinates, and $\epsilon_\|$ and $\epsilon_\bot$ are the anisotropic components of the low-frequency dielectric constant $\varepsilon$. The kinetic energy parts of the electron and hole Hamiltonians (${\hat H}_{e}$, ${\hat H}_{h}$) in the quasi-cubic approximation at zero stress can be found in Refs.\ \onlinecite{lambrecht02,rodina01}. Here we concentrate on $\bm k=0$ and  include the effect of a diagonal strain tensor $\varepsilon$ that preserves the hexagonal symmetry in the framework of the linear DP approach \cite{bir74}. Therefore we write: 
\begin{align}
{\hat H}_{e}(0)-{\hat H}_{h}(0)=&E_{\rm g}+D_1^{\rm
  g}\varepsilon_{zz}+D_2^{\rm g}(\varepsilon_{xx}+\varepsilon_{yy})
\nonumber \\
\label{eg}
&-\Delta_{\rm cr}[{\hat I}_z^2-1]-\frac{\Delta_{\rm so}^{\|}}{3}[({\hat I}_{z}{\hat \sigma}_{z})-1]\\
&-\frac{\Delta_{\rm so}^{\bot}}{3}({\hat I}_{x}{\hat\sigma}_{x}+{\hat
  I}_{y}{\hat \sigma}_{y}), \nonumber \\
\label{dcr}
\Delta_{\rm cr}=&\Delta_{\rm cr}^0-D_{3}\varepsilon_{zz} - D_{4}(\varepsilon_{xx}+\varepsilon_{yy}), \\
\label{dsopar}
\Delta_{\rm so}^{\|}=&\Delta_{\rm so}^{\|, 0}-D_{7}\varepsilon_{zz} - D_{8}(\varepsilon_{xx}+\varepsilon_{yy}), \\
\label{dsoper}
\Delta_{\rm so}^{\bot}=&\Delta_{\rm so}^{\bot, 0}-D_{9}\varepsilon_{zz} - D_{10}(\varepsilon_{xx}+\varepsilon_{yy}).
\end{align}
Here ${\hat\sigma}_{x,y,z}$ are the Pauli matrices, $\hat I_{x,y,z}$ are the projections of the orbital angular momentum operator of the hole ($I=1$), $E_{\rm g}=E(\Gamma_{1c})-E(\Gamma_{5v})$ is the zero stress band gap energy in the absence of the spin-orbit interaction (cf.\ Table\ \ref{EABC}), $\Delta_{\rm cr}=E(\Gamma_{5v})-E(\Gamma_{1v})$ is the crystal field splitting (cf.\ Table\ \ref{EABC}), $\Delta_{\rm so}^{\|}$ and $\Delta_{\rm so}^{\bot}$ are the anisotropic values of the spin-orbit interaction at zero stress, and $D_1^{\rm g}$, $D_2^{\rm g}$, $D_3$, $D_4$, $D_7$, $D_8$, $D_9$ and $D_{10}$ are the constants of the DPs that describe the effects of the external stress which preserves the wurtzite structure of ZnO.

The Hamiltonian given in Eq. (\ref{eg}) results in the well known expressions for the VB splittings\cite{lambrecht02} $E_{\rm AB}=E(\Gamma_{7v}^+)-E(\Gamma_{9v})$ and
$E_{\rm BC}=E(\Gamma_{9v})-E(\Gamma_{7v}^{-})$ with
\begin{align}
-E_{\rm AB}=&\frac{1}{2}\left(\Delta_{\rm so}^{\|} + \Delta_{\rm cr}
-\sqrt{(\Delta_{\rm cr}-\Delta_{\rm so}^{\|})^2+8(\Delta_{\rm so}^{\bot}/3)^2} \right), \\
E_{BC}=&\frac{1}{2}\left(\Delta_{\rm so}^{\|}+\Delta_{\rm cr}
+\sqrt{(\Delta_{\rm cr}-\Delta_{\rm so}^{\|})^2+8(\Delta_{\rm so}^{\bot}/3)^2} \right).
\end{align}
The values of these VB splittings are listed in Table\ \ref{EABC} together with their uniaxial and biaxial strain rates $d_u$ and $d_b$ and their corresponding stress rates $dE/dP$. The first two columns in Table\ \ref{EABC} correlate the terminology of the QP energies from the \emph{ab-initio} calculations and the parameters of the ${\bm k} \cdot {\bm p}$ Hamiltonian, Eq.\ \eqref{eg}. According to this Hamiltonian, the band gap energy between the $\Gamma_9$ VB and the conduction band should be the same as $E_{\rm g}$ without SOC. The values in Table\ \ref{EABC} for $E_{\rm g}$ without SOC are only slightly different from those for $E_{\rm g}(\mathrm{B})$ with SOC. This justifies the application of Eqs.\ \eqref{eg}\,--\,\eqref{dsoper} together with the energies from the \emph{ab-initio} calculations. For reasons of consistency and simplicity, we use the values of $E_{\rm g}(\mathrm{B})$ instead of $E_{\rm g}$ in the Hamiltonian.

From the VB splittings $E_{\rm AB}$, $E_{\rm BC}$, and $\Delta_{\rm cr}$ (cf.\ Table\ \ref{EABC}) we obtain the parameters of the anisotropic spin-orbit interaction as
\begin{align}
\Delta_{\rm so}^\|&=-E_{\rm AB}+E_{\rm BC}-\Delta_{\rm cr}, \\
\Delta_{\rm so}^\bot&=\frac{3}{2\sqrt{2}}
\sqrt{(E_{\rm AB}+E_{\rm BC})^2-(\Delta_{\rm cr}-\Delta_{\rm so}^\|/3)^2}.
\end{align}
For small strain values (corresponding to stress values between $-0.1$~GPa and 0.1~GPa) the spin-orbit constants depend linearly on the strain. The respective zero stress values as well as the strain and stress rates are given in Table \ref{EABC}, respectively.

The DP constants $D_3$ and $D_4$ have the same meaning as in Ref.\ \onlinecite{wrzesinski97}, the constants $D_7$, $D_8$, $D_9$ and
$D_{10}$ describe the stress dependence of the spin-orbit interaction. The constants $D_1^{\rm g}$ and $D_2^{\rm g}$ describe the stress dependence of the band gap $E_{\rm g}$ and are different from the constants $D_1$ and $D_2$ used in Ref.\ \onlinecite{wrzesinski97} because of different definitions of the band gap. The relation between them is given by
\begin{align}
\label{D1g}
D_1&=D_{1}^{\rm g}-D_{3}-D_{7}/3,\\
\label{D2g}
D_2&=D_{2}^{\rm g}-D_{4}-D_{8}/3.
\end{align}

The DP constants for the band gap, $D_{1}^{\rm g}$ and $D_{2}^{\rm g}$, can be determined from the strain rates for $E_{\rm g}(\mathrm{B})$, $d_u$ and $d_b$, as given in Table \ref{EABC} by using
\begin{align}
\label{D1}
D_1^{\rm g}&=\frac{d_u+\nu d_b}{1-\nu R_b},\\
\label{D2}
D_2^{\rm g}&=\frac{R_b d_u+d_b}{2(1-\nu R_b)},
\end{align}
as obtained from Eqs.\ \eqref{du}, \ \eqref{db} with $D_i=D_1^{\rm g}$ and $D_j=D_2^{\rm g}$.

The DP constants $D_3$, $D_4$, $D_{7}$, $D_{8}$, $D_{9}$, and $D_{10}$ are obtained using
\begin{align}
\label{Di}
D_i&= -\frac{d_u+\nu d_b}{1-\nu R_b}\\
\label{Dj}
D_j&= -\frac{R_b d_u+d_b}{2(1-\nu R_b)}
\end{align}
with $i=3,7,9$ and $j=4,8,10$ and the respective strain rates $d_u$ and $d_b$ given in Table\ \ref{EABC}. The crystal-field and spin-orbit splittings are obtained by using $i=3$ and $j=4$ for $\Delta_{\rm cr}$, $i=7$ and $j=8$ for $\Delta_{\rm so}^\|$, and $i=9$ and $j=10$ for $\Delta_{\rm so}^\bot$. Finally, the constants $D_1$ and $D_2$ are calculated based on Eqs.\ \eqref{D1g} and \eqref{D2g}.

\begin{table}
\caption{\label{tEDP}Excitonic deformation potentials for stress which preserves the wurtzite symmetry of ZnO. $D_1$\,--\,$D_4$ are the DPs without spin-orbit interaction, $D_7$\,--\,$D_{10}$ represent the DPs including the spin-orbit coupling. Numbers in parenthesis represent error bars.}
\begin{ruledtabular}
\begin{tabular}{lcccc}
       & This work & Ref.\ \onlinecite{yan12} & Ref.\ \onlinecite{wrzesinski97} & Ref.\ \onlinecite{langer70}\\
       &  (eV)		 &					(eV)						&  			(eV)										& 				(eV) 						\\
\hline
$D_1$ 		& $-3.41$ (0.56) & $-3.06$ & $-3.90$ & $-3.80$\\
$D_2$ 		& $-4.33$ (0.45) & $-2.46$ & $-4.13$ & $-3.80$\\
$D_3$ 		& $-2.26$ (0.36) & $-0.47$ & $-1.15$ & $-0.80$\\
$D_4$ 		& $1.49$ (0.31)  & $0.84$ &  $1.22$ &  $1.40$\\
$D_7$ 		& $-0.37$ (0.04) & -- & -- & --\\
$D_8$ 		& $-0.13$ (0.02) & -- & -- & --\\
$D_9$ 		& $0.09$ (0.02) & -- & -- & --\\
$D_{10}$  & $-0.03$ (0.01) & -- & -- & --\\
\end{tabular}
\end{ruledtabular}
\end{table}

The resulting values for the DP constants were determined using the reported values for $\nu=0.31$ and $R^b=0.93$ by Schleife et al.\cite{Schleife:2007} and are listed in Table\ \ref{tEDP}. The errors of the DPs in parenthesis are estimated assuming errors of 5\% for the strain rates and 10\% for the values of the Poisson ratio and biaxial relaxation coefficient using standard error propagation rules. The experimental determination of specific deformation potentials $D_i$ requires the measurement of the pressure coefficients for at least two independent uniaxial stress directions or the combination of uniaxial and hydrostatic pressure as is expressed by eq. (\ref{def}-\ref{db}) and (\ref{D1}-\ref{Dj}). Experimental data obtained by such combinations of measurements were reported by Langer et al.\cite{langer70} and Wrzesinski et al.\cite{wrzesinski97} The derived values in these works are compared to our calculated deformation potentials in Table\ \ref{tEDP}. The best agreement of our \emph{ab-initio} calculations is obtained with the values in Ref.\ [\onlinecite{wrzesinski97}]. One can see from the data in Table \ref{EABC} that the parameters of the spin-orbit interaction depend on both, uniaxial and biaxial stress. However, the influence of stress on the SOC is found to be much smaller than the influence of stress on the crystal field splitting.

The solution of the Hamiltonian in Eq.\ \eqref{exciton} yields the transition energies $E_{\rm A}$, $E_{\rm B}$, and $E_{\rm C}$ of the A, B, and C excitons for $\bm k=0$ via
\begin{align}
E_{\rm A}&=E_{\rm g}(\mathrm{A})-E_{\rm A}^{b,n},\\
E_{\rm B}&=E_{\rm g}(\mathrm{B})-E_{\rm B}^{b,n} =E_{\rm g}(\mathrm{A})+E_{\rm AB}-E_{\rm B}^{b,n} ,\\
E_{\rm C}&=E_{\rm g}(\mathrm{C})-E_{\rm C}^{b,n}=E_{\rm g}(\mathrm{A})+E_{\rm AB}+E_{\rm AC}-E_{\rm C}^{b,n},
\end{align}
where $E_{\rm A,B,C}^{b,n}$ are the binding energies of the respective excitons of S-symmetry with principle quantum number $n$. The strain and stress dependences of the band gap energies
\begin{align}
E_{\rm g}(\mathrm{A})&=E(\Gamma_{7c})-E(\Gamma_{7v}^{-}),\\
E_{\rm g}(\mathrm{B})&=E(\Gamma_{7c})-E(\Gamma_{9v}),\\
E_{\rm g}(\mathrm{C})&=E(\Gamma_{7c})-E(\Gamma_{7v}^{+})
\end{align}
are directly obtained from the \emph{ab-initio} results and are listed in Table \ref{EABC}. The strain and stress dependences of the exciton binding energies can be determined by the stress dependences of the VB splittings, of the dielectric constants $\epsilon_\|$ and $\epsilon_\bot$, and of the electron and hole effective masses. A theoretical analysis of the exciton binding energies as function of strain and stress is therefore very challenging and beyond the scope of the present work.

The theory of the exciton binding energies in ZnO on the basis of Eq. \ref{exciton} was developed in Ref. \onlinecite{lambrecht02}. It was shown that the binding energies of the 1S ($n=1$ ) and 2S ($n=2$) exciton states are given by $E_{v}^{b,n}=({R_{v}}/{n^2}+\Delta E_{cor}^{v,n})$, where $R_v$ ($v=$ A, B, C) are the effective Rydberg constants of the A, B, and C excitons. $\Delta E_{cor}^{v,n}$ includes the anisotropy corrections, interband interaction corrections, and polaron corrections and accounts for the deviation of about 30 \% from the simple hydrogen-like approximation ${R_{v}}/{n^2}$. The binding energies of the excited exciton states of S-symmetry can be obtained with an accuracy of about 15\% by the following approximation
\begin{align}
\label{Eb}
E_{v}^{b,n}=\frac{E_{v}^{b,1}}{n^2},
\end{align}
where $E_{v}^{b,1}$ ($v=$ A, B, C) are the binding energies of the 1S exciton ground states. In the next section we estimate the A and B exciton binding energies and their stress dependences by analyzing the experimental data using Eq. \ref{Eb}. It should be noted that the experimentally observed transition energies are furthermore affected by the electron-hole exchange interaction and polariton effects which are not included in the Hamiltonian of Eq.\ \eqref{exciton}. We will discuss these effects and their stress dependences in detail in Sec.\ \ref{pol}.

\section{\label{results}Experimental results}

\begin{figure}
\begin{center}
\includegraphics[width=\columnwidth, keepaspectratio]{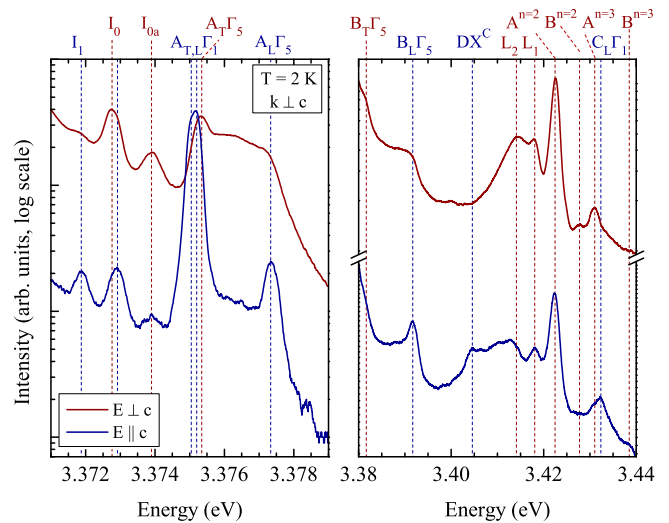}
\caption{\label{fx-polarized}(Color online) Linear polarized PL spectra of a ZnO substrate from Tokyo Denpa at 2~K. Upper red line: $\bm E \bot \bm c$, lower blue line: $\bm E \| \bm c$. Light is excited and detected from the edge of the substrate in $\bm k \bot \bm c$ geometry.}
\end{center}
\end{figure}

In this section we present the experimental results and compare them to our calculations. The exciton-polariton emission lines of ZnO substrates from different manufacturers are studied as a function of uniaxial pressure parallel to the $\bm c$ axis. Linear polarized PL measurements are performed and analyzed considering the dipole selection rules that govern exciton transitions in ZnO \cite{thomas60, hopfield60b, huemmer78phd, wrzesinski97phd}. Figure \ref{fx-polarized} displays the low-temperature PL spectra of a ZnO substrate from Tokyo Denpa between 3.371~eV and 3.440~eV for $\bm E \bot \bm c$ and $\bm E \| \bm c$ polarized light. In the low energy range of the spectrum, several ionized bound exciton transitions ($I_1$, $I_0$, and $I_{0a}$) are observed. While $I_1$ and $I_0$ were identified as ionized bound excitons related to Ga and Al impurities \cite{meyer07, wagner09mej},
$I_{0a}$ is not common in ZnO but can also be identified as an ionized bound exciton based on its localization energy, polarization pattern, and intensity correlation to a neutral bound exciton in the energy range of the $I_5$ line \cite{wagner12ionized}. The significantly higher intensity in the $\bm E \bot \bm c$ configuration demonstrates that most of the bound excitons have a preferred polarization of $\bm E \bot \bm c$ which reflects the larger oscillator strength of the A free exciton in this configuration. The rather pronounced peak in the $\bm E \| \bm c$ configuration in the range of the $I_1$ transition was already reported by Thomas~\cite{thomas60} as well as Loose \emph{et al.} \cite{loose76}

At higher energies above 3.375~eV, the exciton-polaritons of the A, B, and C VBs are clearly resolved. While the longitudinal excitons $\mathrm{A_L}(\Gamma_5$) and $\mathrm{B_L}(\Gamma_5$) can only be observed in the $\bm E \| \bm c$ spectra\cite{ZnOBook}, the transitions from the transverse upper polariton branch have the same energy at $\bm{K}=0$ in the $\bm E \bot \bm c$ configuration, thus, they are visible for both polarizations.\cite{cobet10} The exciton-polaritons of the transverse lower polariton branch $\mathrm{A_T}(\Gamma_5)$ and $\mathrm{B_T}(\Gamma_5)$ are only allowed for $\bm E \bot \bm c$ polarization as seen in Fig.\ \ref{fx-polarized}.

In addition, the transitions of the A($\Gamma_1$) exciton-polaritons are observed in the $\bm E \| \bm c$ spectra. In the case of the C excitons, the two transverse polariton branches of the C($\Gamma_1$) exciton C$_{\rm T}$ and C$_{\rm L}$ are active in the $\bm E \| \bm c$ configuration. The precise determination of the energy of $\mathrm{C_T}$ is only possible by a detailed line shape analysis and careful fitting procedure due to the small oscillator strength of the C($\Gamma_1$) transitions and the superposition of this transition with the $n=2$ excited states of the A excitons. The additional small peaks on the high energy side of the $\mathrm{A}^{n=2}$ lines can be identified as higher excited states of the A and B excitons. Furthermore, three additional emission lines of unknown origin in the energy range between the B and C exciton are observed which will be discussed below. The energy values of the observed optical transitions together with the LT splittings of the A, B, and C exciton-polaritons are listed in Table\ \ref{tZero}.

\begin{table*}
\caption{\label{tZero}Experimental values of zero-stress exciton-polariton energies $E_0$ and uniaxial pressure coefficients $dE/dP_u$ for ground and excited states of a ZnO substrate from Tokyo Denpa. $\omega_{\rm T}$ and $\omega_{\rm L}$ indicate the transverse lower polariton and longitudinal exciton, respectively, $\omega_{\rm L}-\omega_{\rm T}$ is the LT splitting for the exciton-polaritons. Values in brackets indicate the errors in the last digit, precision of absolute energies is limited by the spectral resolution.}
\begin{ruledtabular}
\begin{tabular}{lcccccccccc}
Exciton  & \multicolumn{2}{c}{$\omega_{\rm T}$} & \multicolumn{2}{c}{$\omega_{\rm L}$} & \multicolumn{2}{c}{$n=2$} & \multicolumn{2}{c}{$n=3$} & \multicolumn{2}{c}{$\omega_{\rm L}-\omega_{\rm T}$}\\
  &  $E_0$	&	$dE/dP_u$  &  $E_0$	  &	$dE/dP_u$  &  $E_0$	&	$dE/dP_u$   &  $E_0$	&	$dE/dP_u$  &  $E_0$	&	$dE/dP_u$\\  
  & (eV)    & (meV/GPa)  & (eV)    & (meV/GPa) &   (eV)    & (meV/GPa)  &   (eV)   & (meV/GPa) & (meV)  & (meV/GPa)\\
\hline
$\mathrm{A}(\Gamma_1)$ & 3.37504 & 20.56(5)					& 3.37523 & 	 20.60(5) 	 & --$^a$    &   --$^a$    &   --$^a$  &    --$^a$    & 0.19 & 0.04(7)\\
$\mathrm{A}(\Gamma_5)$ & 3.37538 & 20.4(3)     			& 3.37735 &    21.0(3)    & 3.4226     &  22.7(4)       & 3.4312 &    23.1(6)   & 1.97 & 0.6(4)\\
$\mathrm{B}(\Gamma_5)$ & 3.3815  & 21.7(4)     			& 3.3917  &    22.3(4)    & 3.4280     &  23.9(3)       & 3.4385 &    --$^a$    & 10.2 & 0.6(5)\\
$\mathrm{C}(\Gamma_1)$ & 3.4213  & --$^b$     			& 3.4323  &    13.5(5)    &  --$^a$    & --$^a$      &	 --$^a$	 &    --$^a$   	& 11.0 & --$^b$\\
\hline
$\mathrm{DX^C}$      & 3.4048  & 13.5(14) & & & & & & &\\
$\mathrm{L_2}$		   & 3.4135  & 34.9(20) & & & & & & &\\
$\mathrm{L_1}$		   & 3.4182  & 22.2(4)  & & & & & & &\\
\end{tabular}
\end{ruledtabular}
$^a$ no excited states observed.\\
$^b$ no reliable determination possible due to overlap with $\mathrm{A}^{n=2}$ state.\\
\end{table*}

\begin{figure}
\begin{center}
\includegraphics[width=\columnwidth, keepaspectratio]{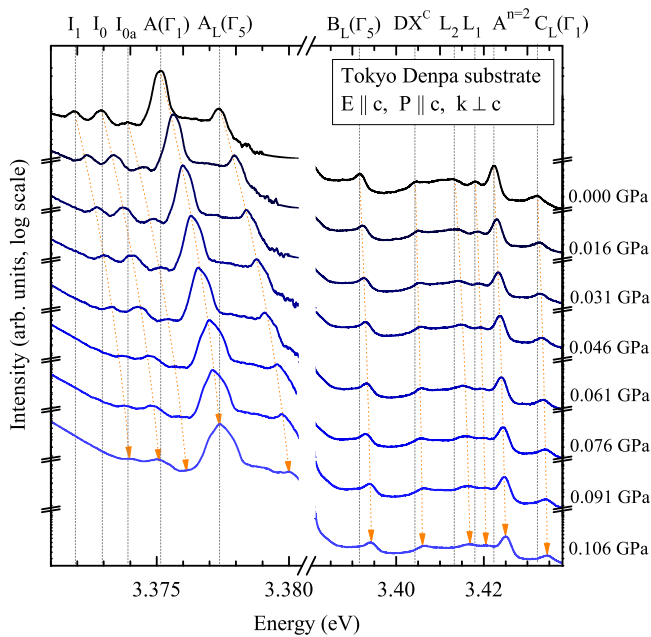}
\caption{\label{fx-parallel}(Color online) Linear polarized PL spectra ($\bm E \| \bm c$) of the exciton-polaritons in ZnO for uniaxial pressure $\bm P \| \bm c$ between 0~GPa and 0.106~GPa at T=2~K.}
\end{center}
\end{figure}

\begin{figure}
\begin{center}
\includegraphics[width=\columnwidth, keepaspectratio]{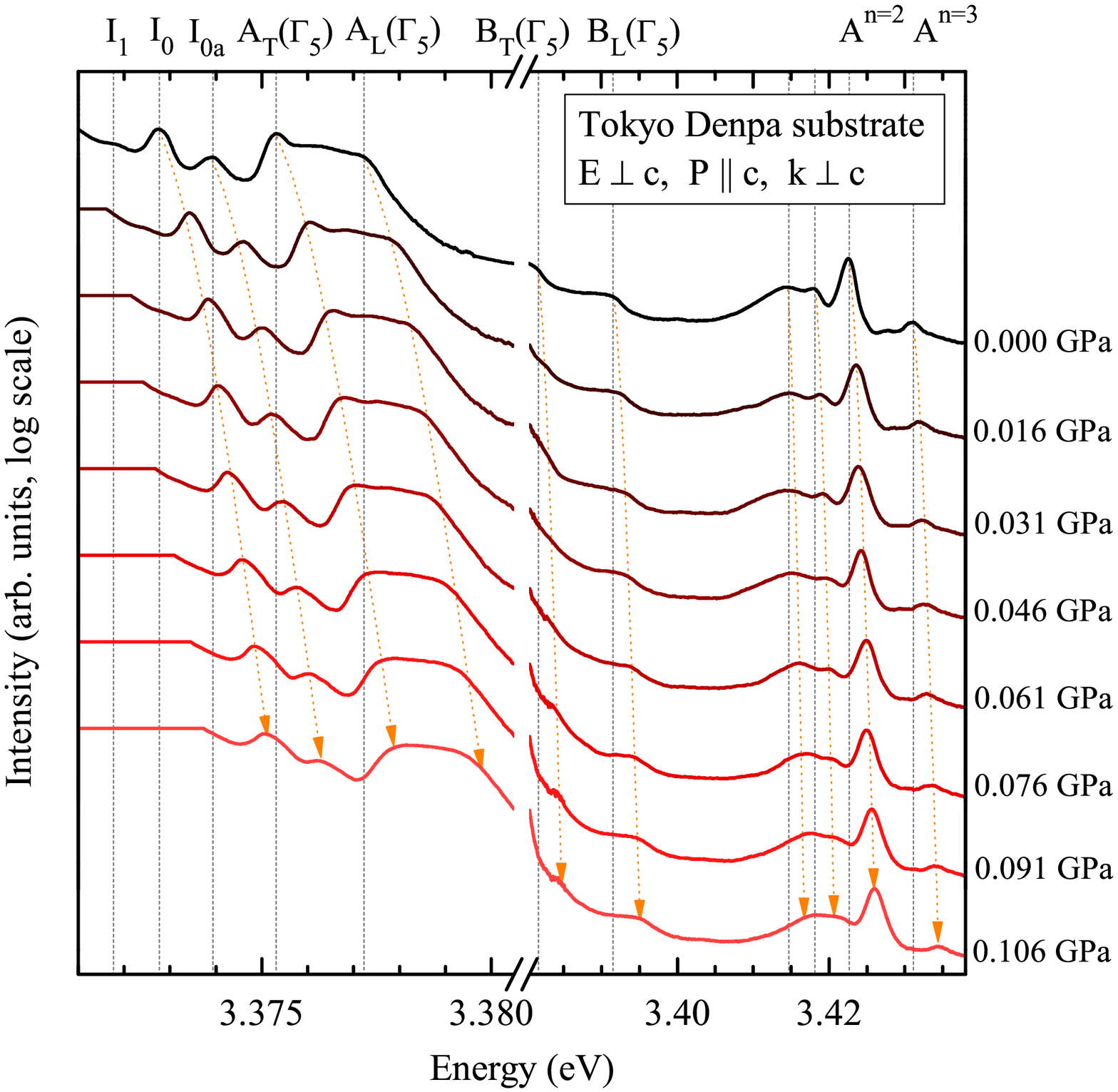}
\caption{\label{fx-perp}(Color online) Linear polarized PL spectra ($\bm E \bot \bm c$) of the exciton-polaritons in ZnO for uniaxial pressure $\bm P \| \bm c$ between 0~GPa and 0.106~GPa at T=2~K.}
\end{center}
\end{figure}

The polarized PL spectra in the range of the exciton-polaritons for uniaxial pressures between 0 and 0.106~GPa are displayed in Figs.\ \ref{fx-parallel} and \ref{fx-perp} for the $\bm E \| \bm c$ and $\bm E \bot \bm c$ configuration, respectively. With increasing uniaxial pressure, a clear shift of all emission lines to higher energies is observed. In addition, specific emission bands such as the A($\Gamma_1$) transition in $\bm E \| \bm c$ polarization reveal an asymmetric broadening with increasing uniaxial pressure. This broadening accounts for the different uniaxial pressure coefficients of the overlapping LT split lines. These pressure coefficients are listed together with the absolute transition energies at zero pressure in Table\ \ref{tZero}. 

Regarding the three unknown emission lines between 3.40~eV and 3.42~eV, the analysis of their polarization pattern and uniaxial pressure coefficients may provide some indications as to their origin. While the 3.4048~eV line is polarized with $\bm E \| \bm c$, the other two transitions can be observed in both polarization directions. Interestingly, the three lines exhibit very different uniaxial pressure coefficients. The peak at 3.4048~eV shifts with a rate of 13.5~meV/GPa which matches the pressure coefficients of the $\mathrm{C_L}(\Gamma_1)$ peak. This agreement strongly suggests a correlation between this emission line and a C VB state which is furthermore supported by the $\bm E \| \bm c$ polarization, thus excluding possible interpretations of this transition as excited states of A or B excitons. In addition, we want to point out that a strong second harmonic generation (SHG) line at comparable energy (3.407~eV) was recently observed which exhibits characteristic properties of a C exciton.\cite{lafrentz13} One possible explanation for the here observed emission line could be a donor bound exciton involving a hole from the C VB band ($\mathrm{DX^C}$). These kind of excited states for donor bound excitons with $\mathrm{B}$ VB holes were studied in detail by Meyer \emph{et al.}\cite{meyer10} and could be observed for a variety of different impurity bound excitons. This interpretation is supported by a localization energy of $E_{loc}=16.5$~meV with regard to the $\mathrm{C_T}$ transition at 3.4213~eV which is in good agreement with typical localization energies of bound excitons in ZnO~\cite{meyer07}. Based on these three arguments, we tentatively identify the 3.4048~eV line as $\mathrm{DX^C}$ transition, possible related to the strongest Al ($I_6$) and Ga ($I_8$) donor bound excitons. The identification of the other two emission lines at 3.4135~eV ($L_2$) and 3.4182~eV ($L_1$) is more difficult. The pressure coefficient of 22.2~meV/GPa might indicate a correlation of the 3.4182~eV transition with an excited state of the A or B exciton, however, the $n=2$ and $n=3$ states of both excitons are clearly identified at higher energies (see Table\ \ref{tZero}). In the case of the 3.4135~eV line, an unusually high shifting rate of 34.9~meV/GPa for uniaxial pressure is observed. At this point, the origin of this transition remains unclear. 

\begin{figure}
\begin{center}
\includegraphics[width=7cm, keepaspectratio]{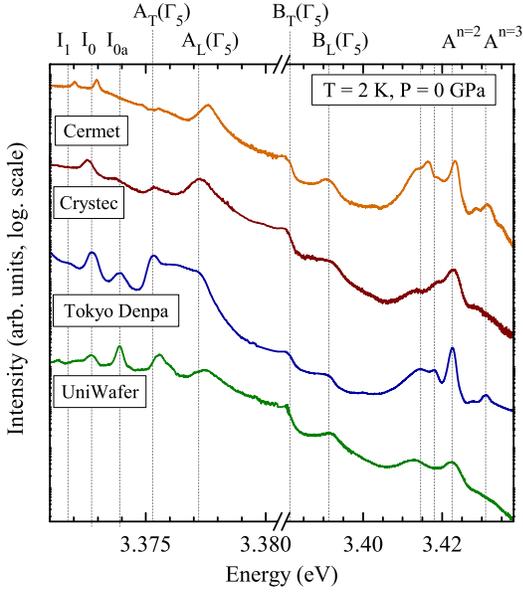}
\caption{\label{plallfx}(Color online) Unpolarized PL spectra of four ZnO substrates in the range of the free exciton-polarion emission lines at T=2~K. Spectra are vertically shifted for clarity.}
\end{center}
\end{figure}

In the following discussion we will focus on the pressure dependence of the fundamental exciton-polariton transitions from the A, B, and C valence bands. In order to enable conclusions of general validity, we compare the absolute energy values and uniaxial pressure coefficients of the exciton-polariton transitions in the sample from Tokyo Denpa to those of other samples from different manufacturers. The low temperature photoluminescence spectra of ZnO substrates from four different suppliers (Cermet, Crystec, Tokyo Denpa, Uniwafer) without external pressure are displayed in Fig.~\ref{plallfx}. The energy positions of the emission lines of bound excitons, free exciton-polaritons, and excited states are indicated for the Tokyo Denpa sample by vertical drop lines. While the main excitonic features are present in all four samples, slight variations in the spectral position, line width, and relative intensity account for the different structural quality and impurity concentration in the various samples. 

\begin{figure}
\begin{center}
\includegraphics[width=7.5cm, keepaspectratio]{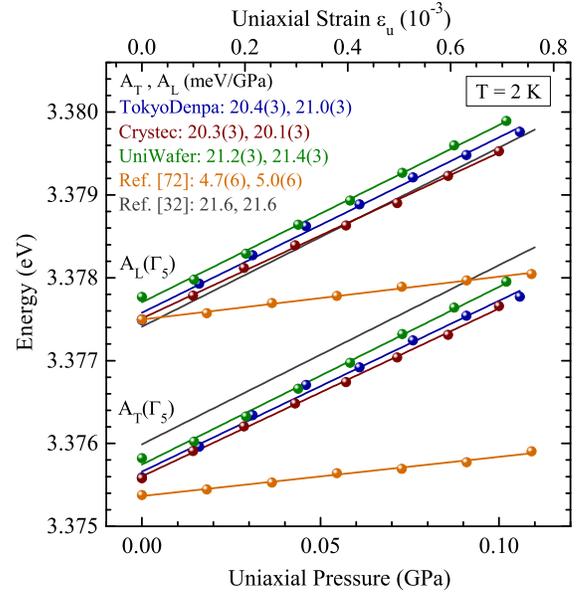}
\caption{\label{uniaxial-shift-all}(Color online) Energy shift of A free exciton-polaritons in different ZnO substrates as function of uniaxial pressure at T=2K.}
\end{center}
\end{figure}

Fig.~\ref{uniaxial-shift-all} displays the energy shift of the $A_T(\Gamma_5)$ and $A_L(\Gamma_5)$ exciton-polariton emission lines as function of uniaxial pressure along the c-axis of the different substrates. The obtained results in this work are compared to our previously reported stress rates\cite{wagner11} for the Cermet substrate and earlier reports by Wrzesinski et al.\cite{wrzesinski97} using two-photon and three-photon spectroscopy under uniaxial stress. The here reported stress rates are found to be in very good agreement with the results of Ref. \onlinecite{wrzesinski97} but are significantly larger than for the Cermet substrate. A possible explanation could be given by the fact that the Cermet sample exhibits a large quantity of extended structural defects as demonstrated by the presence of pronounced luminescence lines related to bound excitons at these centers ($Y$-lines)~\cite{wagner11}. The presence of structural defects might lead to a movement of dislocation (slip) or facilitate the formation of cracks and microcrystals under uniaxial stress, thus effectively releasing strain and reducing the observed shift rates. Furthermore, one should consider that an intrinsic error in the absolute pressure coefficients might occur due to the use of above bandedge excitation which typically generates the PL signal near the surface of the sample. This might lead to deviations in the measured pressure coefficients due to surface defects and inhomogeneities near the edge of the samples. This effect is avoided in e.g.\ the work of Wrzesinski \emph{et al.}\cite{wrzesinski97} who used two and three photon nonlinear spectroscopy and are thus obtaining signal from the entire volume of the crystal instead of just from the surface layer. However, the excellent agreement between the experimentally determined uniaxial pressure coefficients in different ZnO substrates, the results of our theoretical calculations, and the previously reported values using nonlinear spectroscopy underlines the reliability of the here reported values and confirms that the absolute values in Ref.\ \onlinecite{wagner11} are not representative for ZnO in general but are significantly reduced due to the discussed effects. This should also be kept in mind regarding the reported values in the recently published comprehensive study of the 3.324~eV bound exciton line in ZnO by Cullen et al.\cite{cullen13} who used our previously reported stress rates in Ref.\ \onlinecite{wagner11} as calibration for their stress values.

\begin{figure}
\begin{center}
\includegraphics[width=7cm, keepaspectratio]{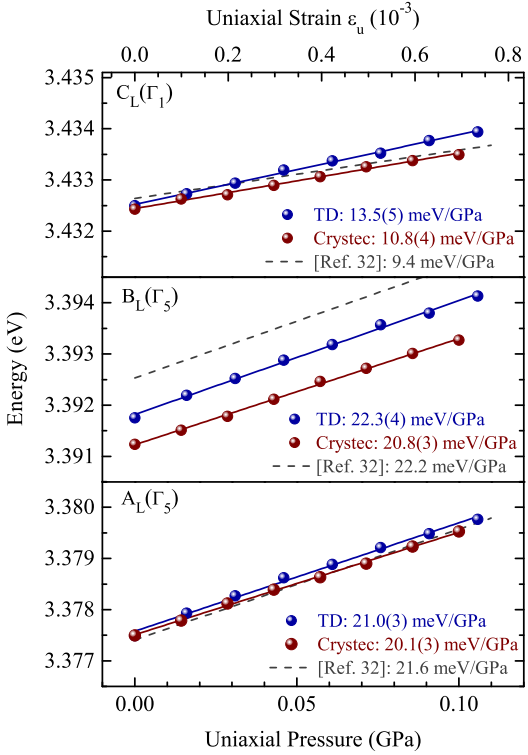}
\caption{\label{uniaxial-shift}(Color online) Energy shift of free exciton-polaritons in different ZnO substrates as function of uniaxial pressure. Dashed gray line: Experimental results reported in Ref.\ \onlinecite{wrzesinski97}.}
\end{center}
\end{figure}

The uniaxial pressure coefficients for the different A, B, and C exciton-polariton transitions are obtained from linear fits to the experimental data in Fig.\ \ref{uniaxial-shift}. These values are summarized in Table~\ref{tUniaxial} together with earlier reported results from Langer \emph{et al.} \cite{langer70} and Wrzesinski \emph{et al.} \cite{wrzesinski97} Apparently, the uniaxial pressure coefficients obtained from our measurements and from literature reports are very consistent with values of 20.1 to 21.6~meV/GPa and 20.8 to 22.9~meV/GPa for the A and B exciton-polaritons, respectively. For the C exciton-polaritons, the variations of the uniaxial pressure coefficients in the different samples are found to be larger with values between 7.7~meV/GPa and 13.5~meV/GPa but still in reasonable agreement with literature reports (cf.\ Table~\ref{tUniaxial}). 

\begin{table*}
\caption{\label{tUniaxial}Zero-stress energies $E_0$ and uniaxial pressure coefficients $dE/dP_u$ of exciton-polariton lines for different ZnO substrates in comparison with values reported in the literature. Values in brackets indicate the errors in the last digit, precision of absolute energies is limited by the spectral resolution.}
\begin{ruledtabular}
\begin{tabular}{lcccccccc}
           & \multicolumn{2}{c}{$\mathrm{A_T}(\Gamma_5)$} & \multicolumn{2}{c}{$\mathrm{A_L}(\Gamma_5)$} & \multicolumn{2}{c}{$\mathrm{B_L}(\Gamma_5)$} & \multicolumn{2}{c}{$\mathrm{C_L}(\Gamma_1)$}\\
Sample & $E_0$ & $dE$/$dP_u$ & $E_0$ & $dE$/$dP_u$ & $E_0$ & $dE$/$dP_u$ & $E_0$ & $dE$/$dP_u$\\
                                    & (eV)      & (meV/GPa) & (eV)  & (meV/GPa)&     (eV)  & (meV/GPa)& (eV)     & (meV/GPa)\\
\hline
Tokyo Denpa                         & 3.37538  	& 20.4(3) & 3.37735  & 21.0(3) & 3.3917    & 22.3(4)  & 3.4323    & 13.5(5) \\
CrysTec                             & 3.3756    & 20.3(3) & 3.3775   & 20.1(3) & 3.3912    & 20.8(3) 	& 3.4324    & 10.8(4) \\
UniWafer                            & 3.3758    & 21.2(3) & 3.3777   & 21.4(3) & 3.3914    & 22.9(3) 	& 3.4318    & 7.7(6) \\ \hline
Ref.\ \onlinecite{wrzesinski97}     & 3.37599 	& 21.6 & 3.37741  & 21.6 & 3.39253    	& 22.2  & 3.43264 	& 9.4 \\
Ref.\ \onlinecite{langer70}         &   --      &  --    & 3.378    & 20.8 & 3.389     & 21.3  & 3.429     & 10.1 \\
\end{tabular}
\end{ruledtabular}
\end{table*}

\begin{figure}
\begin{center}
\includegraphics[width=6.5cm, keepaspectratio]{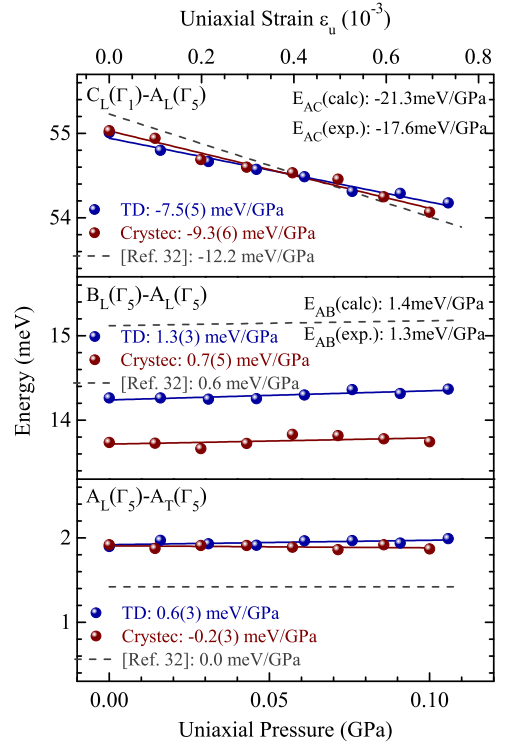}
\caption{\label{uniaxial-splitting}(Color online) Splitting between the exciton-polaritons involving hole states from the A, B, and C VBs as function of uniaxial pressure. Dashed gray line: Experimental results according to Ref.\ \onlinecite{wrzesinski97}.}
\end{center}
\end{figure}

The energies and stress rates of the observed transitions are not expected to be identical with the theoretical energies $E_{\rm A}$, $E_{\rm B}$, and $E_{\rm C}$ and their shift rates described in Secs.\ \ref{stress} and \ref{defpot} since the contributions of the exchange interaction and the exciton-polariton coupling have to be included in the theoretical description. This is discussed in detail in section \ref{pol}. Nevertheless, the measured energies of $\mathrm{A_T}$ and $\mathrm{B_T}$ can be used as good approximation to obtain the band energies $E_{\rm A}$ and $E_{\rm B}$ and their stress rates. Based on the observation of the $n=2$ and $n=3$ transitions of the A and B excitons, the exciton binding energies at zero stress can be estimated using the hydrogen-like model described in Eq.\ \eqref{Eb}, leading to $E_{\rm A}^{b,n=1}  \approx 62.5$ meV and $E_{\rm B}^{b,n=1}  \approx 61.6$ meV. The uniaxial stress rates for the binding energies of the A and B excitons are found to be equal with a value of $2.8$ meV/GPa.

For the different energy splittings in the Tokyo Denpa sample we obtain experimental stress rates of $dE_{\rm g}(\mathrm{A})/P_u = 23.4$ meV/GPa and $dE_{\rm g}(\mathrm{B})/P_u = 24.7$ meV/GPa. These values are only slightly smaller than the theoretical results of $26.4$ meV/GPa and $27.8$ meV/GPa for the A and B band-gap energies (see Table \ref{EABC}). The shift rate of the $\mathrm{A_T}-\mathrm{B_T}$ exciton splitting of $dE_{\rm AB}/P_u =1.3$ meV/GPa is in excellent agreement with the calculated value of 1.4 meV/GPa as listed in Table~\ref{EABC}.

Alternatively, one can use the $\mathrm{A_L}(\Gamma_5)$, $\mathrm{B_L}(\Gamma_5)$, and $\mathrm{C_L}(\Gamma_1)$ energies (cf.\ Fig.\ \ref{uniaxial-shift}) as reference to estimate the VB splittings as function of uniaxial pressure $dE_{\rm AB}/P_u$ and $dE_{\rm AC}/P_u$ as shown in Fig.\ \ref{uniaxial-splitting}. Using the $\mathrm{A_L}(\Gamma_5)-\mathrm{B_L}(\Gamma_5)$ splitting to estimate the stress rates $dE_{\rm AB}/P_u$ for the three different ZnO samples we obtain experimental values of 0.7~meV/GPa (Crystec), 1.3~meV/GPa (Tokyo Denpa), and 1.5~meV/GPa (UniWafer) which are in good agreement with the theoretical prediction of 1.4 meV/GPa (see Table \ref{EABC}). In contrast, the $\mathrm{A_L}(\Gamma_5)-\mathrm{C_L}(\Gamma_1)$ splitting as function of uniaxial pressure yields stress rates between $-7.5$ meV/GPa and $-13.7$ meV/GPa for the three different samples which is significantly smaller than the theoretical result of $dE_{\rm AC}/P_u=-21.3$ meV/GPa (see Table \ref{EABC}). A possible reason could be that the theoretical description above does neither account for the differences in the stress rates of the binding energies of the A, B, and C excitons nor for the effects of stress on the exchange interaction and the LT splittings. These effects are including in the theoretical considerations in the next section. For the A and B excitons the stress rates are found to be equal, however, the data for the stress rate of the binding energy of the C exciton is missing. For the following analysis this rate is assumed to be equal to those of the A and B excitons which seems reasonable due to the reported similarity in the case of hydrostatic pressure.\cite{mang95} The LT splitting of the A exciton-polaritons as function of uniaxial pressure is displayed in the lower panel of Fig.\ \ref{uniaxial-splitting}; the stress dependence is very small. Similar considerations apply for the LT splitting of the B exciton-polariton (cf.\ Table \ref{tZero}). In the case of the C exciton-polaritons, an experimental determination of the stress rate of the LT splitting is not possible due to the previously discussed overlap of the $\mathrm{C_T}$ line with the A$^{n=2}$ lines. Thus, the pressure dependence of the LT splitting of the C exciton has to be obtained from theoretical considerations.

\section{\label{pol}Effect of stress on the exchange interaction and exciton-polariton fine structure}

In this section, we discuss the effect of stress on the exchange interaction and LT splitting of the exciton-polaritons by deriving the analytical expression for the stress dependent exciton and exciton-polariton energies. We use the experimentally observed uniaxial stress rates for the exciton-polariton lines in order to compute the missing stress rates for the $\mathrm{C_T}$ line as well as for the $E_{\rm A}$, $E_{\rm B}$, and $E_{\rm C}$ exciton energies in order to extract the experimental stress rates for the energy splittings and compare them with the results of the \emph{ab-initio} calculations in Table \ref{EABC}.

The excitonic Hamiltonian of Eq.\ \eqref{exciton} does not include the effects of the electron-hole exchange interaction which contains short range and long range contributions. The short range exchange interaction can be added as\cite{lambrecht02,thang85}
\begin{equation}
\hat H_{\rm exch} = \frac{\kappa}{2} (1- \hat {\bm \sigma}_e\hat
{\bm \sigma}_h ) \, ,
\end{equation}
where $\kappa$ is the characteristic energy of the isotropic short-range exchange interaction and $\hat {\bm \sigma}_{e}$ and $\hat {\bm \sigma}_{h}$ are the electron and hole spin operators, respectively. The energy $\kappa$ is defined according to Ref. \onlinecite{elliotbook} by $\kappa=\Omega |\phi_{1S}(0)|^2 J$, where $\Omega$ is the volume of the unit cell, $\phi_{1S}({\bm r_e}-{\bm r_h})$ is the 1S hydrogen-like function which describes the relative motion of electron and hole, and $J$ is an atomic-like exchange energy. The use of the same characteristic energy $\kappa$ for the A, B, and C excitons is justified since their binding energies and their corresponding $\phi_{1S}$ functions are very similar.\cite{lambrecht02, thang85}

In the case of well separated noninteracting A, B, and C excitons, the Hamiltonian $\hat H_{\rm ex} + \hat H_{\rm exch}$ yields the energies of the $\mathrm{A_T}(\Gamma_5)$, $\mathrm{B_T}(\Gamma_5)$, and $\mathrm{C_T}(\Gamma_5)$ transitions as $E_{\rm A} +\kappa a^2$, $E_{\rm B} +\kappa$, and $E_{\rm C} +\kappa b^2$, respectively. The values $a$ and $b$ describe the contributions of the $z$ component to the basis function of the $\Gamma_{7v}^{+}(\mathrm{A})$ and $\Gamma_{7v}^{-}(\mathrm{C})$ valence bands so that $a^2+b^2=1$. The explicit expressions for them can be found in Refs.\ \onlinecite{lambrecht02,rodina01}.

In ZnO, the value of $a^2$ is very close to 1 and $b^2$ is small. Since the exchange interaction energy between the A and B excitons is comparable with the energy separation $E_{\rm AB}$, this interaction needs to be taken into account. The transverse exciton states with $\Gamma_5$ symmetry are then given by:
\begin{align}
\label{eq1}
\mathrm{A_T}=&\frac{1}{2} \Big( E_{\rm A} + E_{\rm B}+ \kappa (a^2 +1)\\
\nonumber
&-\sqrt{(E_B-E_{\rm A} + \kappa b^2)^2 + 4 \kappa^2 a^2} \Big),\\
\label{eq2}
\mathrm{B_T}=&\frac{1}{2} \Big( E_{\rm A} + E_{\rm B}+ \kappa (a^2 +1) \\
\nonumber
&+\sqrt{(E_{\rm B}-E_{\rm A} + \kappa b^2)^2 + 4 \kappa^2 a^2} \Big).
\end{align}

For the A and C excitons of $\Gamma_1$ symmetry one obtains $\mathrm{A_{T1}} \equiv \mathrm{A_T}(\Gamma_1)=E_{\rm A}+ 2\kappa b^2$ and $\mathrm{C_T}(\Gamma_1)=E_{\rm C}+ 2\kappa a^2$, without taking an interaction between them into account. Although the oscillator strength for $\mathrm{A_T}(\Gamma_1)$ is expected to be small in ZnO (proportional to $b^2$), the $\mathrm{A_T}(\Gamma_1)$ transition is stronger than the $\mathrm{C_T}(\Gamma_1)$ transition in the $\bm E \| \bm c$ spectrum due to the intraband relaxation and larger population of holes to the A VB at low temperatures. In contrast to the A-B energy splitting, the energy separation between the A and C excitons in ZnO is much larger than the
characteristic exchange energy (cf.\ Tables \ref{EABC} and \ref{tZeroParameter}). Nevertheless, it is important to include the exchange interaction between the A and C excitons for a proper description of the stress effects on the exciton-polariton fine structure since the $E_{\rm AC}$ energy is strongly affected by stress. Including this interaction, we obtain the following expressions for the energies of the A($\Gamma_1$) and C($\Gamma_1$) excitons:
\begin{align}
\mathrm{A_{T1}}= E_{\rm A} + 2\kappa b^2 - \frac{4\kappa^2 a^2b^2}{E_{\rm C}-E_{\rm A}}
\, ,  \label{eq3} \\
\mathrm{C_T}=E_{\rm C} + 2\kappa a^2 + \frac{4\kappa^2 a^2b^2}{E_{\rm C}-E_{\rm A}}
\, . \label{eq4}
\end{align}

In the used ${\bm k} \bot \bm c$ geometry (we may assume that ${\bm k}$ is directed e.g.\ along the $y$ direction), the $\Gamma_{5x}$ exciton states (for each A and B exciton) are the transverse states with the energies $\mathrm{A_T}$ and $\mathrm{B_T}$, respectively, while the $\Gamma_{5y}$ states are the longitudinal excitons with energies $\mathrm{A_L}$ and $\mathrm{B_L}$, respectively. Their energies can be derived from the condition $\epsilon^{\bot}(\omega) = 0$ where $\epsilon^{\bot}(\omega)$ is the frequency dependent dielectric function for an electric field ${\bm E} \bot \bm c$ with frequencies close to the A and B excitonic resonances.\cite{ivchenko82} Longitudinal excitons can be observed for the ${\bm E} \| \bm c$ polarization. Due to the interaction of the transverse excitons with photons, a lower and an upper polariton branch are formed. Their energies can be derived from the condition $\epsilon^{\bot}(\omega) = (ck/\omega)^2$ with $c$ being the speed of light. As the energies of the upper transverse polariton branches coincide with the energies of the longitudinal excitons at $\bm k=0$, they are often labeled $\mathrm{A_L}$ and $\mathrm{B_L}$, whereas the energies of the lower transverse polariton branch at the $\Gamma$ point are marked as $\mathrm{A_T}$ and $\mathrm{B_T}$.

In order to determine $\epsilon^{\bot}(\omega)$ for frequencies close to the exciton resonances and to obtain the longitudinal-transversal (LT) splitting it is instructive to follow the general approach described by Ivchenko.\cite{ivchenko82} Neglecting the interaction between the A and B excitons and considering them as well separated resonances, one obtains $\epsilon^{\bot}(\omega)=\epsilon_b^{\bot}(1+(2a^2K_\bot A_T)/(A_T^2-\hbar^2\omega^2))$ near the $A_T$ exciton resonance and $\epsilon^{\bot}(\omega)=\epsilon_b^{\bot}(1+(2K_\bot B_T)/(B_T^2-\hbar^2\omega^2))$ close to the energy of the $B_T$ exciton. Here, $K_\bot=4\pi d_\bot^2/\epsilon_b^{\bot}$ is the characteristic energy for the dipole interaction with an electric field ${\bm E} \bot \bm c$, $d_\bot$ are the dipole matrix element between Bloch functions of the $\Gamma_5$ valence band and conduction band, and $\epsilon_b^{\bot}$ is the background (high frequency) dielectric constant. The resulting upper branch energies are given by $A_L\approx A_T + a^2 K_{\bot}$ and $B_L\approx B_T + K_{\bot}$. Apparently, the LT splittings $a^2 K_{\bot}$ for the $\mathrm{A}(\Gamma_5$) exciton and $K_{\bot}$ for the $\mathrm{B}(\Gamma_5$) exciton become almost equal as the parameter $a^2$ is close to 1 for ZnO. The large differences for the LT splittings of the $\mathrm{A}(\Gamma_5$) and $\mathrm{B}(\Gamma_5$) excitons observed in the experiment can be explained by the interaction between themselves. Accounting for the simultaneous contributions of the A($\Gamma_5$) and B($\Gamma_5$) excitons to the dielectric function and their short-range exchange interaction, we derive $\epsilon^{\bot}(\omega)$ following the same general approach:
\begin{align}
\label{ebot}
\frac{\epsilon^{\bot}}{\epsilon_b^{\bot}}&=1+\frac{2K_\bot}{[(\hbar\omega)^2-\mathrm{A_T}^2] [(\hbar\omega)^2-\mathrm{B_T}^2]}\\ 
\nonumber
&\cdot \Big[{\rm A}_{\rm T}{\rm B}_{\rm T}(E_{\rm A}+a^2E_{\rm B})\\
\nonumber
&+(\hbar\omega)^2 [(E_{\rm A}+a^2E_{\rm B})-(\mathrm{A_T}+\mathrm{B_T})(1+a^2)] \Big] .
\end{align}
 Using the conditions $\mathrm{A_L}-\mathrm{A_T} \ll \mathrm{A_T}$ and $\mathrm{B_L}-\mathrm{B_T} \ll \mathrm{B_T}$, we find approximate energies of the longitudinal excitons/upper polariton branches of the A($\Gamma_5$) and B($\Gamma_5$) excitons as:
\begin{align}
\label{eq5} 
\mathrm{A_L}&\approx\frac{\mathrm{A_T}+\mathrm{B_T}}{2} + K_\bot a^2\\
\nonumber
&-\frac{1}{2} \sqrt{(\mathrm{B_T}-\mathrm{A_T})^2+4K_\bot^2a^4+8K_\bot\kappa a^2},\\
\label{eq6}
\mathrm{B_L}&\approx\frac{\mathrm{A_T}+\mathrm{B_T}}{2} + K_\bot a^2 \\
\nonumber
&+\frac{1}{2} \sqrt{(\mathrm{B_T}-\mathrm{A_T}+2K_\bot
b^2)^2+4K_\bot^2(1-2b^2)+8K_\bot\kappa a^2}.
\end{align}

In the case of $\Gamma_1$ states, all excitons in the ${\bm k} \bot \bm c$ geometry are transverse. For ${\bm E} \| \bm c$ polarized light one can observe also upper transverse polariton branches of the A($\Gamma_1$) and C($\Gamma_1$) excitons at the energies $\mathrm{A_{L1}} \equiv \mathrm{A_L}(\Gamma_1)$ and $\mathrm{C_L}$. These energies can be determined from the condition $\epsilon^{\|}= 0$, where $\epsilon^{\|}(\omega)$ is now the frequency dependent dielectric function for the electric field ${\bm E} \| \bm c$ in the range of frequencies close to the A and C excitonic resonances, respectively. Considering again the simultaneous contributions of the A($\Gamma_1$) and C($\Gamma_1$) excitons and their short-range exchange interaction and following the general approach of Ref. \cite{ivchenko82} we derive $\epsilon^{\|}(\omega)$ as:
\begin{align}
\label{epar}
\frac{\epsilon^{\|}}{\epsilon_b^{\|}}&=1+\frac{2K_\|}{[(\hbar\omega)^2-\mathrm{A_{T1}}^2] [(\hbar \omega)^2-\mathrm{C_T}^2]}\\
\nonumber
&\cdot\Big[\mathrm{A_{T1}}\mathrm{C_T}(b^2E_C+a^2E_{\rm A})\\
\nonumber
&+(\hbar\omega)^2 [(b^2E_C+a^2E_{\rm A})-(\mathrm{A_{T1}}+\mathrm{C_T})] \Big],
\end{align}
where $\epsilon_b^{\|}$ is the background (high frequency) dielectric constant, $K_\|=4\pi d_\|^2/\epsilon_b^{\|}$ is the characteristic energy for the dipole interaction with an electric field ${\bm E} \| \bm c$, and $d_\|$ is the dipole matrix element between the Bloch functions of the $\Gamma_1$ valence band and conduction band). Using the conditions $\mathrm{A_{L1}}-\mathrm{A_{T1}} \ll \mathrm{A_{T1}}$ and $\mathrm{C_L}-\mathrm{C_T} \ll \mathrm{C_T}$ as well as $\kappa, K_\| \ll \mathrm{C_T}-\mathrm{C_L}$, we find approximate energies of the longitudinal excitons/upper polariton branches of the A($\Gamma_1$) and C($\Gamma_1$) excitons as: 
\begin{align}
\label{eq7}
\mathrm{A_{L1}} \approx \mathrm{A_{T1}}+ K_\| b^2-  \frac{2K_\|^2b^2}{\mathrm{C_T}-\mathrm{A_{T1}}}-\frac{2K_\| \kappa
a^2b^2(1+2b^2)}{\mathrm{C_T}-\mathrm{A_{T1}}}, \\
\label{eq8} 
\mathrm{C_L} \approx \mathrm{C_T}+ K_\| a^2 + \frac{2K_\|^2a^2}{\mathrm{C_T}-\mathrm{A_{T1}}}+\frac{2K_\| \kappa
a^2b^2(1+2a^2)}{\mathrm{C_T}-\mathrm{A_{T1}}}.
\end{align}
It should be noted that even without considering the interaction between A($\Gamma_1$) and C($\Gamma_1$) excitons, the theory predicts the LT splitting $b^2 K_{\|}$ of the  A($\Gamma_1$) exciton to be much smaller than the LT splitting $a^2 K_{\|}$ of the C($\Gamma_1$) exciton in agreement with the experimental data. However, the inclusion of this interaction is crucial for a proper description of the stress dependences because of the large stress rate for the energy separation between A and C excitons.

\begin{table}
\caption{\label{tZeroParameter}Zero-stress energies $E_0$ and uniaxial pressure coefficients $dE/dP_u$ for the A, B, and C exciton energies, exchange and dipole interaction parameters $\kappa$, $K_\|$, $K_\bot$, and wave function parameter $a$ as derived by solving equations \eqref{eq1}\,--\,\eqref{eq4}, \eqref{eq5}, \eqref{eq6}, \eqref{eq7}, and \eqref{eq8}.}
\begin{ruledtabular}
\begin{tabular}{cccccc}
 &  					& \multicolumn{2}{c}{This Work} & Ref.\ \onlinecite{wrzesinski97phd}	& Ref.\ \onlinecite{fiebig93}\\
 &	Symmetry  & $E_0$ & $dE/dP_u$ & $E_0$ &  $E_0$ \\
 & $\Gamma_i$ & (eV) & (meV/GPa) & (eV) & (eV) \\ 
\hline
\multicolumn{6}{l}{Input parameters}\\
\hline
$\mathrm{A_{T1}}$ 	& $\Gamma_1$  & 3.37504 & 20.56 & 3.37516 	& 3.3756\\
$\mathrm{A_{L1}}$ 	& $\Gamma_1$  & 3.37523 & 20.60 & 3.37539 	&	3.3759\\
$\mathrm{A_T}$ 		& $\Gamma_5$  & 3.37538 & 20.4 & 3.37599 & 3.3759 \\
$\mathrm{A_L}$ 		& $\Gamma_5$  & 3.37735 & 21.0 & 3.37741 & 3.3778 \\
$\mathrm{B_T}$ 		& $\Gamma_5$  & 3.3815  & 21.7 & 3.38256 & 3.3816 \\
$\mathrm{B_L}$ 		& $\Gamma_5$  & 3.3917  & 22.3 & 3.39253 & 3.3929 \\
$\mathrm{C_L}$ 		& $\Gamma_1$  & 3.4323  & 13.5 & 3.43264 & 3.4327 \\
\hline
\multicolumn{6}{l}{Output values}\\
\hline
$\mathrm{C_T}$ 		& $\Gamma_1$  & 3.4213 &  2.6	& 3.42162 & 3.4209 \\
$E_{\rm A}$ 		&             & 3.37501 & 20.6  & 3.37505 & 3.37555 \\
$E_{\rm B}$ 		&             & 3.38106 & 21.9	& 3.38112 & 3.38118 \\
$E_{\rm C}$ 		&             & 3.42051 &  3.0 	& 3.41930 & 3.42015 \\
$\kappa$	&	 					  & 0.00041 & $-0.21$	& 0.00122 & 0.00040 \\
$K_\bot$  &   					& 0.00627 & 0.65 	& 0.00591 & 0.00694 \\
$K_\|$		& 						& 0.00838 & 5.7	 	& 0.00844 & 0.00897 \\
$a^*$    	&             & 0.98112 & -0.0004 & 0.97643 & 0.96960 \\
\end{tabular}
\end{ruledtabular}
$^*$ The parameter $a$ is dimensionless, the units of all other parameters are as specified.\\
\end{table}

Consequently, we have now derived eight equations \eqref{eq1}\,--\,\eqref{eq4}, \eqref{eq5}, \eqref{eq6}, \eqref{eq7}, and \eqref{eq8} which completely determine the exciton-polariton energies $\mathrm{A_{T1}}$, $\mathrm{A_{L1}}$, $\mathrm{A_T}$, $\mathrm{A_L}$, $\mathrm{B_T}$, $\mathrm{B_L}$, $\mathrm{C_T}$, and $\mathrm{C_L}$ as experimentally observed in the polarized PL spectra (see Tables \ref{tZero} and \ref{tZeroParameter}). From this set of equations we can calculate all eight unknown variables. These are the zero stress exciton energies $\mathrm{C_T}$, $E_{\rm A}$, $E_{\rm B}$, and $E_{\rm C}$, the exchange and dipole interaction parameters $\kappa$, $K_\|$, $K_\bot$, and the wave function parameter $a$. The resulting values are listed in Table \ref{tZeroParameter}. For comparison, we have also calculated these parameters using the exciton-polariton energies reported by Wrzesinski \emph{et al.} \cite{wrzesinski97phd} and Fiebig \emph{et al.}\cite{fiebig93} (cf.\ Table \ref{tZeroParameter}). One can see that the resulting energies and parameters are in very good agreement which justifies the described approach.

In the next step, we use the zero stress energies and the experimentally determined uniaxial stress rates for the lines $\mathrm{A_{T1}}$, $\mathrm{A_{L1}}$, $\mathrm{A_T}$, $\mathrm{A_L}$, $\mathrm{B_T}$, $\mathrm{B_L}$, and $\mathrm{C_L}$ in order to compute the stress dependence of all output parameters by solving equations \eqref{eq1}\,--\,\eqref{eq8}. The resulting stress rates are also listed in Table \ref{tZeroParameter}. It is important to note that the computed results for the $\mathrm{C_T}$ line are very sensitive to the input parameters. Thus, even small errors of the input stress rates or zero stress energies may result in about one order of magnitude larger errors for the position and stress rates of the $\mathrm{C_T}$ line as compared to the input values. The stress rate for the $\mathrm{C_T}$ line of 2.6 meV/GPa is found to be significantly smaller than the measured value of $13.5$ meV/GPa for the $\mathrm{C_L}$ line. This large difference in the stress rates for $\mathrm{C_L}$ and $\mathrm{C_T}$ lines (and thus the large stress rate for the LT splitting of the $\mathrm{C}(\Gamma_1)$ exciton of 10.9 meV/GPa) is partly caused by the stress dependence of the dipole interaction energy $a^2 K_{\|}$ of the $\mathrm{C}(\Gamma_1)$ exciton and partly due to the large stress rate for the energy splitting between A and C excitons. Compared to the C excitons, the stress rate for the LT splitting of the $\mathrm{A}(\Gamma_1)$ excitons is significantly smaller because of the much weaker dipole interaction energy $b^2 K_{\|}$. Nevertheless, the small difference in the stress rates of the $\mathrm{A_T}(\Gamma_1)$ and $\mathrm{A_L}(\Gamma_1)$ exciton lines can still be observed in Fig. 3 as a broadening of the $\mathrm{A}(\Gamma_1)$ exciton line with increasing uniaxial stress.

Using the computed uniaxial stress rates $E_{\rm A}$, $E_{\rm B}$, and $E_{\rm C}$ and assuming the stress rates of the A, B, and C exciton binding energies to be equal (2.8 meV/GPa), we obtain the uniaxial rates $dE/dP_u$ for the energy splittings $E_{\rm g}(\mathrm{A})$, $E_{\rm g}(\mathrm{B})$, $E_{\rm g}(\mathrm{C})$, $E_{\rm AB}$, $E_{\rm BC}$, and $E_{\rm AC}$ listed in the eights column of Table \ref{EABC} as experimentally determined rates. These shifting rates are in a very good agreement with our computed rates by \emph{ab-initio} calculations (sevenths column of Table \ref{EABC}). The procedure described above considerably improves the estimation for the A-C exciton splitting. Indeed, using the rates for $E_{\rm A}$ and $E_{\rm C}$, we obtain a value of about $-17.6$~meV/GPa for the $E_{\rm AC}$ splitting in good agreement with the theoretical result of $-21.3$~meV/GPa. The remaining discrepancy could be caused e.g.\ by the fact that we have assumed the same stress rate for the C exciton binding energy as experimentally determined for the A and B excitons as well as because of the still large uncertainty for the stress rate of the $\mathrm{C_T}$ exciton line.

\section{Conclusions}
\label{conclusion}
We have presented a detailed theoretical and experimental study on the effects of stress on the band gap energies and the exciton-polariton fine structure in wurtzite ZnO. The theoretical approach combined \emph{ab-initio} calculations for the QP energies as function of uniaxial and biaxial stress with the ${\bm k} \cdot {\bm p}$ modeling of the band structure Hamiltonian and the exciton-polariton splittings. These calculations were complemented by polarization dependent luminescence measurements of the exciton-polariton emission lines in different ZnO substrates as function of uniaxial stress. Based on the joint experimental and theoretical approach, clear evidence is found that the ordering of the topmost A($\Gamma_7$) and B($\Gamma_9$) valence bands in ZnO remains unchanged even for large uniaxial and biaxial stress. The zero-stress energies and the strain and stress rates for the band gap energies, valence band splittings, crystal-field splitting, and anisotropic spin-orbit coupling are computed for uniaxial, biaxial, and hydrostatic pressure. While the crystal-field spitting is found to be highly sensitive to uniaxial and biaxial stress, the spin-orbit coupling is almost unaffected by stress, thus explaining the stability of the symmetry ordering of the A and B valence bands in ZnO. In addition, the full set of symmetry preserving deformation potentials is determined including the spin-orbit interaction and its anisotropy.

Based on the combination of polarized luminescence measurements and ${\bm k} \cdot {\bm p}$ modeling, the uniaxial stress rates are determined for the longitudinal-transverse (LT) splitting of the exciton polaritons (including the LT splitting of the A($\Gamma_1$) exciton-polariton) and their excitonic binding energies. While the stress dependence of the LT splitting of the A and B exciton-polaritons is found to be small for uniaxial pressure, the LT splitting of the C exciton-polariton is significant which is explained by the large stress dependences of the dipole interaction energy and the crystal-field interaction. It is shown that only through the inclusion of the stress dependence of the exchange interaction and the LT splitting in the theoretical description, it is possible to achieve an accurate description of the experimental data. Furthermore, we also obtain the zero-stress energies and uniaxial stress rates for the dipole interaction and wave function parameters.

In summary, the combination of theoretical and experimental techniques in this work proved that the valence band ordering in ZnO is unaffected by any realistic amount of strain due to e.g. lattice mismatch and different coefficients of thermal expansion in hetero-epitaxially grown layers or different impurity concentrations in bulk material. Our results also provide a multitude of important electronic parameters including the spin-dependent deformation potentials as well as the energies and stress rates of the crystal-field interaction, spin-orbit coupling, exchange interaction, and exciton-polariton splittings.
 
\begin{acknowledgments}
The research presented here received funding from the European Community's Seventh Framework Program (FP7/2007-2013) under grant agreement No.\ 211956 and from the Deutsche Forschungsgemeinschaft (Project No.\ Be1346/20-1). A.V.R. acknowledges support of the Russian Foundation for Basic Research (Grant No. 13-02-00888-a) and the support received from the Swiss National Science Foundation. A.S. thanks the Carl-Zeiss-Stiftung for support. We acknowledge grants of computer time from the HLRS Stuttgart and the LRZ Munich. Support from the DFG within SFB 787 and the cluster of excellence "UniCat" is acknowledged. Part of this work was performed under the auspices of the U.S.\ Department of Energy at Lawrence Livermore National Laboratory under Contract DE-AC52-07A27344.

\end{acknowledgments}

\end{document}